\documentclass[letterpaper,twocolumn,10pt]{article}
\usepackage{usenix-2020-09}

\usepackage{graphicx}
\usepackage{caption}
\usepackage{subcaption}
\usepackage{rotating}
\usepackage{latexsym}
\usepackage{amssymb}
\usepackage{bbm}
\usepackage{pifont}
\usepackage[nointegrals]{wasysym}
\usepackage{stmaryrd}

\usepackage{epsfig,amsmath,amsfonts,epsfig,multirow,makecell,caption,soul,csquotes,color,wrapfig,subcaption,mathtools,bm,spverbatim,booktabs,tcolorbox,diagbox}
\usepackage[e]{esvect}

\def\tablename{Table}
\def\figurename{Figure}

\usepackage{caption}
\makeatletter
\newif\if@restonecol
\makeatother

\usepackage[boxed, ruled, vlined, linesnumbered]{algorithm2e}
\SetKwRepeat{Do}{do}{while}

\newenvironment{changemargin}[2]{\begin{list}{}{
	\setlength{\topsep}{0pt}\setlength{\leftmargin}{0pt}
	\setlength{\rightmargin}{0pt}
	\setlength{\listparindent}{\parindent}
	\setlength{\itemindent}{\parindent}
	\setlength{\parsep}{0pt plus 1pt}
	\addtolength{\leftmargin}{#1}\addtolength{\rightmargin}{#2}
	}\item}
	{\end{list}}

\usepackage[first=0,last=9]{lcg}
\usepackage{colortbl}
\definecolor{Gray}{gray}{0.8}
\colorlet{Red}{red!10!white}
\colorlet{Blue}{blue!10!white}

\usepackage{hyperref}

\newcommand{\msec}[1]{\S\ref{#1}}
\newcommand{\mref}[1]{\,\ref{#1}}
\newcommand{\meq}[1]{Eq.\,\ref{#1}}
\newcommand{\mcite}[1]{\cite{#1}}

\newcommand{\meg}{\textit{e.g.}\xspace}
\newcommand{\mie}{\textit{i.e.}\xspace}

\newcommand{\mcounter}[1]{(\textit{#1})}

\newtcolorbox{mtbox}[1]{left=0.25mm, right=0.25mm, top=0.25mm, bottom=0.25mm, sharp corners, colframe=red!50!black, boxrule=0.5pt, title={#1}, fonttitle=\bfseries, coltitle=red!50!black, attach title to upper={\ --\ }}

\usepackage{scalerel}[2016/12/29]

\makeatletter
\providecommand{\leadsfrom}{%
  \mathrel{\mathpalette\reflect@squig\relax}%
}
\newcommand{\reflect@squig}[2]{%
  \reflectbox{$\m@th#1\leadsto$}%
}
\makeatother

\newcommand{\dnns}{{\small DNNs}\xspace}
\newcommand{\ml}{{\small ML}\xspace}

\def\eqref#1{equation~\ref{#1}}
\def\1{\bm{1}}

\DeclareMathAlphabet{\mathsfit}{\encodingdefault}{\sfdefault}{m}{sl}
\SetMathAlphabet{\mathsfit}{bold}{\encodingdefault}{\sfdefault}{bx}{n}

\def\gE{{\mathcal{E}}}
\def\gF{{\mathcal{F}}}
\def\gG{{\mathcal{G}}}

\def\gK{{\mathcal{K}}}
\def\gL{{\mathcal{L}}}

\def\gN{{\mathcal{N}}}

\def\gQ{{\mathcal{Q}}}
\def\gR{{\mathcal{R}}}
\def\gS{{\mathcal{S}}}

\def\gV{{\mathcal{V}}}

\begin{document}

\newcommand{\kg}{{\small KG}\xspace}
\newcommand{\kgs}{{\small KGs}\xspace}

\newcommand{\qa}{{\sc QA}\xspace}
\newcommand{\cve}{{\small CVE}\xspace}
\newcommand{\cves}{{\small CVEs}\xspace}
\newcommand{\rec}{{\sc Rec}\xspace}
\newcommand{\krl}{{\small KRL}\xspace}
\newcommand{\mrr}{{\small MRR}\xspace}
\newcommand{\hit}{{\small HIT@$K$}\xspace}
\newcommand{\hito}{{\small HIT@$1$}\xspace}

\newcommand{\ndcg}{{\small NDCG@$K$}\xspace}
\newcommand{\ndcgf}{{\small NDCG@5}\xspace}

\newcommand{\akp}{{\sc Roar}$_\mathrm{kp}$\xspace}
\newcommand{\aqp}{{\sc Roar}$_\mathrm{qp}$\xspace}
\newcommand{\aco}{{\sc Roar}$_\mathrm{co}$\xspace}

\newcommand{\system}{{\sc Roar}\xspace}

\newcommand{\NA}{----}
\newcommand{\kgp}{{\sc Kgp}\xspace} % KG poisoning
\newcommand{\lqe}{{\sc Lqe}\xspace} % Evasion
\newcommand{\cop}{{\sc CoP}\xspace} % Collaborative
\newcommand{\ota}{{\small OTA}\xspace} % optimization & projection
\newcommand{\vect}[1]{\boldsymbol{#1}} % boldface-italic math 

\newcommand{\ting}[1]{\textcolor{purple}{[#1]}\xspace}

\newcommand{\ren}[1]{\textcolor{orange}{[#1]}\xspace}

\title{\Large Towards Robust Reasoning over Knowledge Graphs}
% \author{Anonymous}
\author{
{\rm Zhaohan Xi}$^\dagger$ \,\, {\rm Ren Pang}$^\dagger$ \,\, {\rm Changjiang Li}$^\dagger$ \,\, {\rm Shouling Ji}$^\ddagger$ \,\, {\rm Xiapu Luo}$^\ast$ \,\, {\rm Xusheng Xiao}$^\S$ \,\, {\rm Ting Wang}$^\dagger$\\
$^\dagger$Pennsylvania State University \quad
$^\ast$The Hong Kong Polytechnic University \\
$^\ddagger$Zhejiang University and Ant Financial  \quad
$^\S$ Case Western Reserve University \\
} % end author

\maketitle
\thispagestyle{empty}

\begin{abstract}
Answering complex logical queries over large-scale knowledge graphs (\kgs) represents an important artificial intelligence task, entailing a range of applications. Recently, knowledge representation learning (\krl) has emerged as the state-of-the-art approach, wherein \kg entities and the query are embedded into a latent space such that entities that answer the query are embedded close to the query. Yet, despite its surging popularity, the potential security risks of \krl are largely unexplored, which is concerning, given the increasing use of such capabilities in security-critical domains (\meg, cyber-security and healthcare).
 
This work represents a solid initial step towards bridging this gap. We systematize the potential security threats to \krl according to the underlying attack vectors (\meg, knowledge poisoning and query perturbation) and the adversary's background knowledge. More importantly, we present \system\footnote{\system: \underline{R}easoning \underline{O}ver \underline{A}dversarial \underline{R}epresentations.}, a new class of attacks that instantiate a variety of such threats. We demonstrate the practicality of \system in two representative use cases (\mie, cyber-threat hunting and drug repurposing). For instance, \system attains over 99\% attack success rate in misleading the threat intelligence engine to give pre-defined answers for target queries, yet without any impact on non-target ones. Further, we discuss potential countermeasures against \system, including filtering of poisoning facts and robust training with adversarial queries, which leads to several promising research directions.
\end{abstract}

\section{Introduction}
\label{sec:intro}

\begin{figure}[!t]
    \centering
    \epsfig{file = 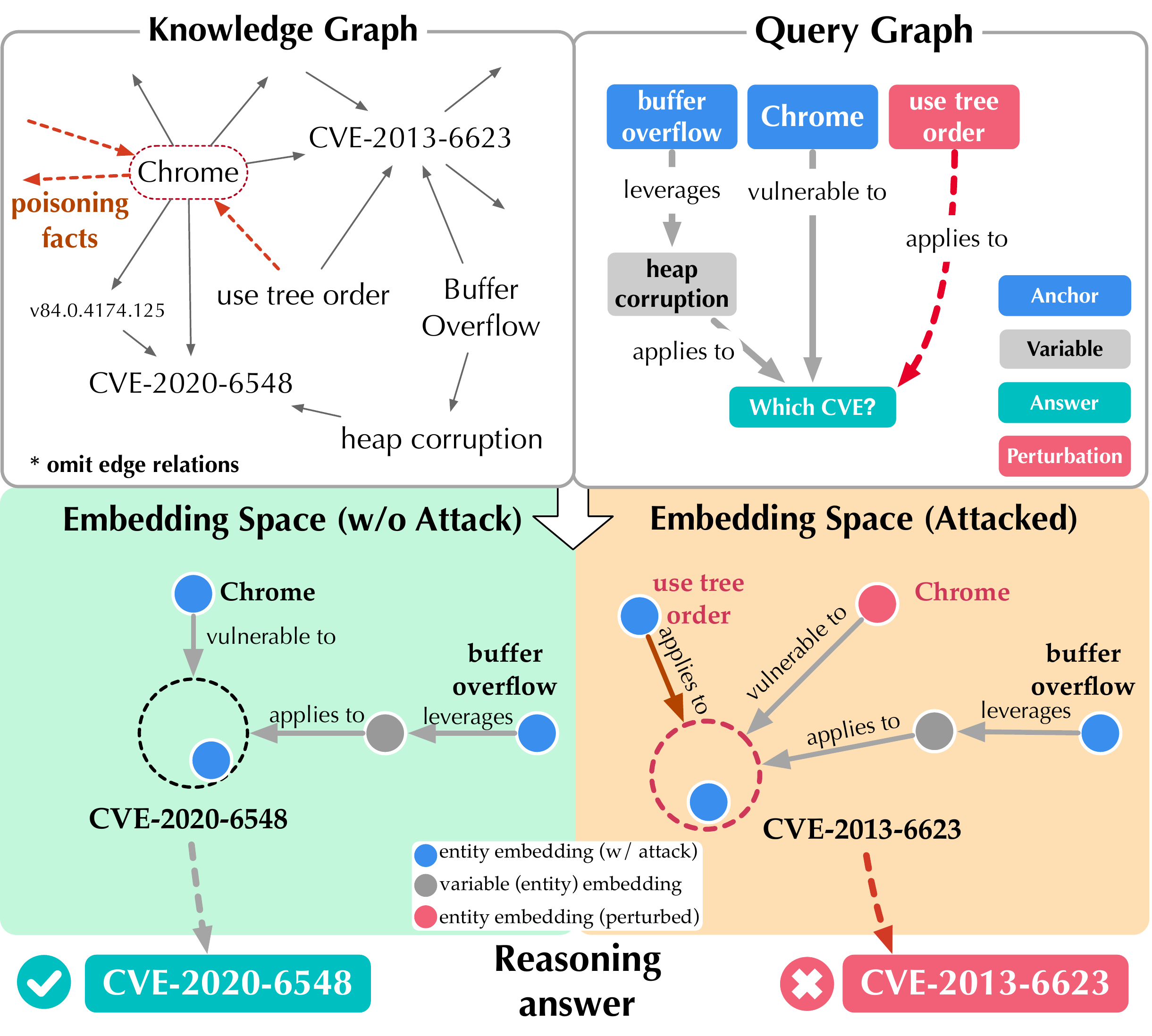, width = 84mm}
    \caption{Illustration of \krl.}
    \label{fig:example}
\end{figure}

A knowledge graph (\kg) is a structured representation of human knowledge about ``facts'', with entities, relations, and descriptions respectively capturing real-world objects (or abstract concepts), their relationships, and their semantic properties. Answering complex logical queries over \kgs represents an important artificial intelligence task, entailing a range of applications. For instance, in Figure\mref{fig:example}, the security analyst queries for the most likely vulnerability (\cve) that is being exploited, based on observations regarding the incident (\meg, attack technique, tactics, and affected product). 

Typically, the computational complexity of processing such queries grows exponentially with the query size\mcite{logic-query-embedding}, which hinders its use over large \kgs. Recently, {\em knowledge representation learning} (\krl) has emerged as a promising approach, wherein \kg entities and a query are projected into a latent space such that the entities that answer the query are embedded close to each other. As shown in Figure\mref{fig:example}, \krl reduces answering an arbitrary query to simply identifying entities with embeddings most similar to the query, thereby implicitly imputing missing relations\mcite{missing-relation} and scaling up to large-scale \kgs in various domains\mcite{yago,gkg,makg}.

Surprisingly, in contrast to the intensive research on improving the capabilities of \krl, its security risks are largely unexplored. This is highly concerning given \mcounter{i} the importance of integrating domain knowledge in artificial intelligence tasks has been widely recognized\mcite{domain-knowledge}; \mcounter{ii} \kgs have emerged as one of the most popular representations of domain knowledge in various security-sensitive domains (\meg, cyber-security\mcite{mittal2019cyber} and healthcare\mcite{zhu2020knowledge}); and \mcounter{iii} \krl has become the state-of-the-art approach to process complex queries over such \kgs\mcite{query2box,logic-query-embedding}. We thus wonder: 
%In this paper, we seek to bridge this gap by answering the following research questions:

\vspace{1pt}
RQ$_1$ -- {\em What are the potential security threats to \krl?}

\vspace{1pt}
RQ$_2$ -- {\em How effective are the attacks in practical settings?} 

\vspace{1pt}
RQ$_3$ -- {\em Are there any countermeasures against such attacks?} 

\vspace{2pt}
{\bf  Our work --} This work represents a solid initial step towards answering these questions. 

\vspace{1pt}
RA$_1$ -- First, we characterize the potential security threats to \krl. As illustrated in Figure\mref{fig:example}, the adversary may disrupt the reasoning of \krl through two attack vectors: \mcounter{i} knowledge poisoning -- polluting the \kgs by committing poisoning facts, and \mcounter{ii} query perturbation -- modifying the queries by adding misleading variables/relations. We create a threat taxonomy according to the underlying attack vectors as well as the adversary's background knowledge.

\vspace{1pt}
RA$_2$ -- Further, we present \system, a new class of attacks that instantiate the above threats. We evaluate the practicality of \system attacks in two representative use cases, cyber-threat hunting and drug repurposing. It is empirically demonstrated that \system is highly effective against the state-of-the-art \krl systems in both domains. For instance, in both cases, \system attains over 99\% attack success rates in misleading \krl to give pre-defined answers for target queries, yet without any impact on non-target queries. 

\vspace{1pt}
RA$_3$ - Finally, we discuss potential countermeasures and their technical challenges. According to the attack vectors, we consider two mitigation strategies: \mcounter{i} filtering of poisoning \kg facts and \mcounter{ii} robust training with adversarial queries. We reveal that there exists an interesting synergy between the two defenses and also a delicate trade-off between \krl performance and attack resilience.
%which points to a few promising research directions.

\vspace{2pt}
{\bf  Contributions --} To the best of our knowledge, the work represents the first in-depth study on the security of \krl. Our contributions are summarized as follows. 

We characterize the potential security threats to \krl and reveal the design spectrum for the adversary with varying capability and knowledge. 

We present \system, a new class of attacks that instantiate various threats to \krl, which highlights with the following features: \mcounter{i} it leverages both knowledge poisoning and query perturbation as the attack vectors; \mcounter{ii} it is highly effective and evasive; \mcounter{iii} it assumes limited knowledge regarding the target \krl system; and \mcounter{iv} it realizes both targeted and untargeted attacks. 

We discuss potential mitigation, which sheds light on improving the current practice of training and using \krl, pointing to several promising research directions.

% \begin{figure*}
%     \centering
%     \epsfig{file = figures/illustration_attack.pdf, width = 180mm}
%     \caption{\footnotesize .}
%     \label{fig:illustration:attack}
% \end{figure*}
\section{Preliminaries}
\label{sec:background}

We first introduce fundamental concepts and assumptions used throughout the paper. %The important notations are summarized in Table\mref{tab:notations}.

% \subsection{Reasoning over Knowledge Graphs} 

\vspace{2pt}
{\bf Knowledge Graph --} A knowledge graph (\kg) $\gG = (\gN, \gE)$ consists of nodes $\gN$ and edges $\gE$ connecting them. Each node $v \in \gN$ represents an entity and each edge $v \xrightarrow{r} v' \in \gE$ indicates that there exists relation $r \in \gR$ (where $\gR$ is a finite set of relation types) from $v$ to $v'$. In other words, $\gG$ comprises a set of {\em facts} $\{\langle v, r, v' \rangle \}$ with $v, v' \in  \gN$ and $v \xrightarrow{r} v' \in  \gE$.

\vspace{2pt}
{\bf Query --} A variety of reasoning tasks can be performed over \kgs. Here, we focus on two types of queries.

\vspace{1pt}
{\em Entity Query --} In an entity query, one asks for a specific entity that satisfies given logical constraints, which are often defined using first-order conjunctive logic with existential ($\exists$) and conjunction ($\wedge$) operations. Formally, letting $v_?$ be the entity of interest, $\gK_q$ be the set of known {\em anchors}, $\gV_q$ be a set of existentially quantified {\em variables}, and $\gE_q$ be a set of relations, an entity query $q \triangleq (v_?, \gK_q, \gV_q, \gE_q)$ is defined as:
\begin{equation}
\begin{split}
    & q[v_?] = v_? \,.\, \exists \gV_q : \wedge_{v \xrightarrow{r} v' \in \gE_q}  v \xrightarrow{r} v'  \\
    & \text{s.t.} \;\, v \xrightarrow{r} v' = \left\{
    \begin{array}{l}
    v \in \gK_q, v' \in \gV_q \cup \{v_?\}, r\in \gR  \\
     v, v' \in \gV_q \cup \{v_?\}, r\in \gR
    \end{array}
    \right.
\end{split}
\end{equation}
As an example, in Figure\mref{fig:example}, $v_?$ is the sink (\textsf{\small CVE}), \textsf{\small Chrome} is an anchor, while \textsf{\small heap corruption} is a variable.  

Intuitively, $q$ can be represented as a {\em dependency graph}, of which the nodes and edges correspond to the entities and relations in $q$, respectively. We say that a query $q$ is valid if its dependency graph is directed acyclic with $\gK_a$ as the source nodes and $v_?$ as the unique sink node\mcite{logic-query-embedding,query2box}. 

\vspace{1pt}
{\em Relation Query --} In a relation query, one asks for the potential relation, which is currently not present in the \kg, between two given entities $v, v'$. Formally, a relation query $q \triangleq (r_?, v, v')$ is defined as: \begin{equation}
    q[r_?] = r_? \,.\, \exists v \xrightarrow{r_?} v'
\end{equation}
To determine $r_?$, it often requires to account for the context surrounding $v$ and $v'$ in the \kg. 

%For instance, \ting{add an example}

\vspace{2pt}
{\bf Knowledge Representation Learning --} Recently, knowledge representation learning (\krl) has emerged as the state-of-the-art approach to process such queries. It projects \kgs and query $q$ to a latent space, such that entities that answer $q$ are embedded close to $q$.

\begin{figure}[!t]
    \centering
    \epsfig{file = 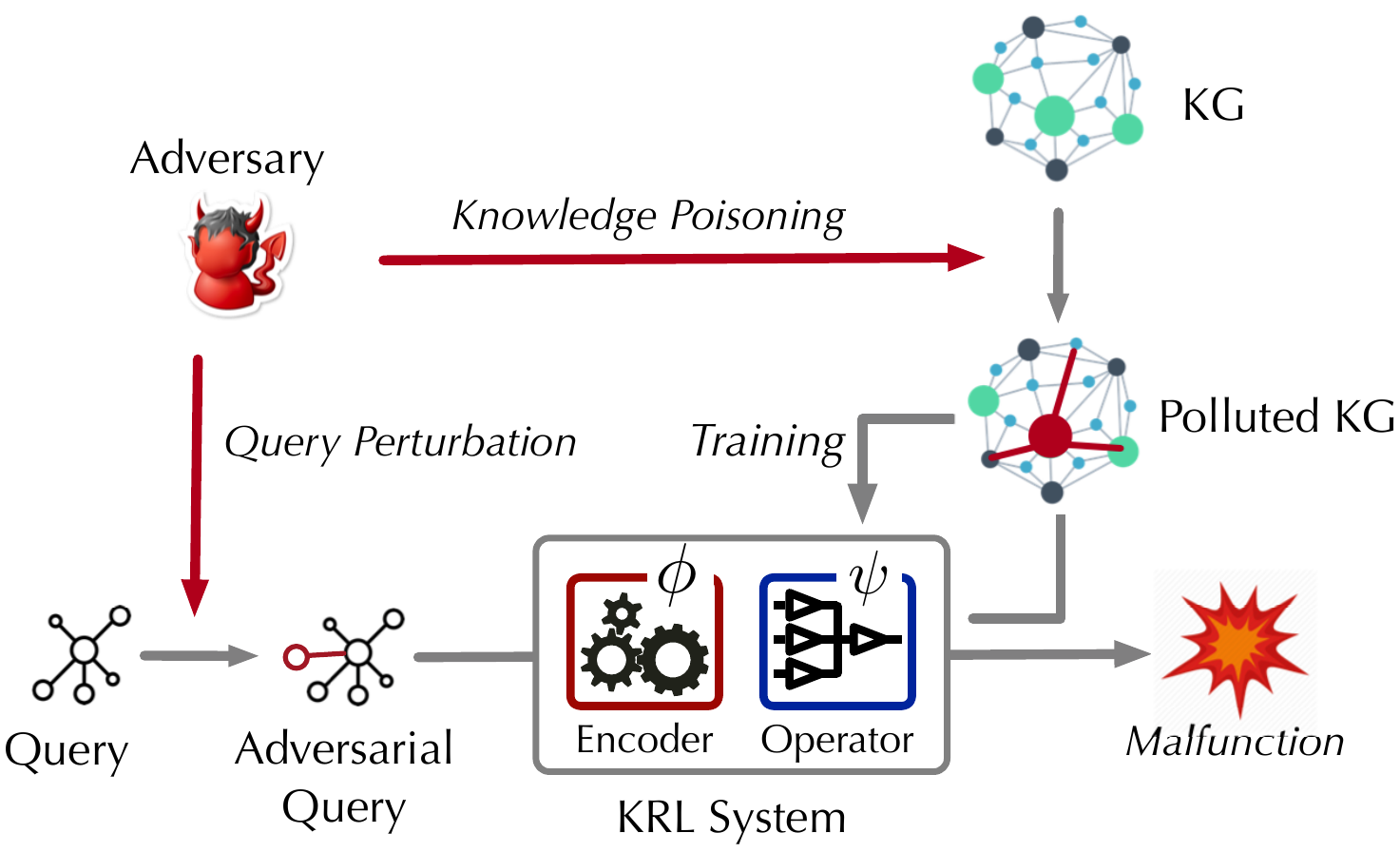, width = 84mm}
    \caption{Illustration of \system attacks against \krl.}
    \label{fig:pipeline}
\end{figure}

As shown in Figure\mref{fig:pipeline}, a \krl system $\gamma$ typically comprises two components, {\em encoder} $\phi$ that projects entity $v$ to its embedding $\phi_v$, and relation $r$-specific {\em operator} $\psi_r$ that 
performs geometric transformation to $\phi_v$ to compute entities with relation $r$ to $v$. 
\krl often samples a set of query-answer pairs from the \kg as the training set and trains $\phi$ and $\psi$ in a supervised manner.

% measures the ``fitness'' of relation $r$ between two embeddings $\phi_v$ and $\phi_{v'}$. 
% A plethora of embedding designs have been proposed, including vectors\mcite{kg-traversal}, boxes\mcite{query2box}, and distributions\mcite{beta-embedding}.

\vspace{1pt}
{\em Entity Query --} To process entity query $q$, one may derive $q$'s {\em computation graph} from its dependency graph with a set of operators: \mcounter{i} projection - given entity set $\gS$, it computes the entities with relation $r$ to $\gS$; \mcounter{ii} intersection - given a set of entity sets $\{\gS_1, \gS_2, \ldots, \gS_n \}$, it computes $\cap_{i=1}^n \gS_i$. Intuitively, the computation graph specifies the procedure to answer $q$ as traversing the \kg: starting from the anchors, it iteratively applies the two oprators, until reaching the unique sink\mcite{kg-traversal}. 
% The training objective is thus to learn the optimal geometric transformation of projection and intersection operators.

%{\em Entity Query --} To process entity queries, one may derive the {\em computation graph} from $q$'s dependency graph with a set of operators: \mcounter{i} projection - given a set of entities $\gS \subseteq \gN$ and relation $r$, it computes the entities in $\gN$ with relation $r$ to $\gS$, that is $\cup_{v \in \gS }\gA_r(v)$, where $\gA_r(v) = \{v'  \in \gN \,|\, \exists v \xrightarrow{r} v' \} $; \mcounter{ii} intersection - given a set of entity sets $\{\gS_1, \gS_2, \ldots, \gS_n \}$, it computes $\cap_{i=1}^n \gS_i$. Intuitively, the computation graph specifies the reasoning procedure to answer $q$ as traversing $\gG$: starting from the anchor entities, it iteratively applies the two operations, until reaching the unique sink entity\mcite{kg-traversal}. The training objective is thus to learn the optimal geometric transformation of projection and intersection operators.

% To answer entity queries, \krl project \kg $\gG$ and query $q$ into the latent space such that entities that answer $q$ are embedded close to $q$ (typically measured using $L_2$ distance). Under this setting, answering an arbitrary query is reduced to simply identifying entities nearest to the query embedding in the latent space, which implicitly imputes missing relations\mcite{missing-relation} and scales up to large \kgs and queries. 

% More specifically, 

\vspace{1pt}
{\em Relation Query --} To answer relation queries, \krl projects the \kg to the latent space and typically applies graph neural networks (\meg,\mcite{rgcn}) as the link prediction operator that aggregates multi-relational graphs to predict the missing relations. %The training objective is thus to learn the optimal link prediction operator.

% To process a conjunctive query $q$, the straightforward approach is to perform graph matching between $q$'s dependency graph with the KG $\gG$. While simple and intuitive, this approach suffers major drawbacks including: (i) its computational complexity grows exponentially with $q$'s size; and (ii) it operates incorrectly when $q$ has missing relations

\section{Threat Characterization}
\label{sec:threatmodel}

We systematize the threats to \krl according to the underlying attack vectors and the adversary's knowledge. %For simplicity, we exemplify with entity queries to illustrate the attacks. 

\subsection{Adversary model}
\label{ssec:objective}

We first define the adversary model.

\vspace{2pt}
{\bf Adversary's objectives --} Let $\gQ$ be the testing query set and $\gQ^*$ be the set of queries targeted by the adversary ($\gQ^* \subseteq \gQ$). In an {\em untargeted} attack, the adversary aims to force the \krl system to make erroneous reasoning over $\gQ^*$; in a {\em targeted} attack, the adversary attempts to force the system to return a target answer $A$ when processing $\gQ^*$.

In both cases, to be evasive against inspection, the attack must have a limited impact on the system's performance on non-target queries $\gQ \setminus \gQ^*$. 

% reasoning performances on the query set $\ssub{Q}{\backslash tar}$ without $\ssub{\gL}{tar}$ should keep as the original, hence ensuring an ignorable side effect on the victim system.

% A reasoning system leverages first-order logic (FOL) on \kg or queries to infer results. Correspondingly, our attacks are triggered by specific FOL that naturally existed among queries. We use $\ssub{\gL}{tar}$ to denote the targeted FOL in a path format, and use $\ssub{Q}{tar}$ to denote a set of queries with $\ssub{\gL}{tar}$.

% An adversary expects the reasoning on $\ssub{Q}{tar}$ always leads to erroneous results. Specifically, there are two attack objectives: 
% \begin{mitemize}
% \setlength\itemsep{0pt}
% \item {\bf Targeted inference -- }  when processing $\ssub{Q}{tar}$, the reasoning system always returns a specific answer $\ssub{a}{tar}$ denoted by the adversary.
% \item {\bf General degradation --}  reasoning results on $\ssub{Q}{tar}$ are generally wrong, \mie, reasoning performances on $\ssub{Q}{tar}$ is decreasing.
% \end{mitemize}

%\subsection{Adversary's capability}
%\label{ssec:attack:vector}

\vspace{2pt}
{\bf Adversary's capability --} As illustrated in Figure\mref{fig:pipeline}, we consider two different attack vectors.

% Developing a reasoning system involves knowledge embedding and model training, using private training queries with monitored training process; hence, the whole development process is trustworthy. 

% However, noisy information may locate in \kg downloaded from external sources; an adversary can rely on intentionally injected knowledge facts to affect a victim system. Additional facts in \kg influences graph structure hence change the first-order logic (FOL) represented by \kg. Even though it is impractical to perturb a \kg anywhere, an adversary can focus on some entities and perturb corresponding localities with reported facts.

% In addition, developing a reasoning system is agnostic of downstream use cases, leaving spaces for specific evasion. Assuming no system is perfect even though well-developed, an adversary queries the system to find logics that lead to erroneous results, then inject those logics into victim queries. Since a downstream user constructs her queries from the wild (\meg, testing a software product or parsing biomedical reports), an adversary can inject perturbations to data sources in advance with prior knowledge of potential victims.

% Even worse, an adversary with strong accessibility can collaboratively use those two vectors (\mie, \kg and use-case queries) to develop a more effective attack schema. We show how to realize a collaborative attack based on co-optimization in Section\mref{sec:attack}.

\vspace{1pt}
{\em Knowledge Poisoning --} Before training, the adversary may commit poisoning facts which are then integrated by \krl to construct the \kgs. We argue that this attack vector is practical: most \kgs are automatically built using public, crowdsourced information (\meg,\mcite{wiki}); due to the massive amount of information, it is often challenging to conduct thorough fact checking\mcite{kg-incompleteness}, which opens the door for knowledge poisoning. To make the attack evasive, we require that the number of poisoning facts is limited.

\vspace{1pt}
{\em Query Perturbation --} At inference, the adversary may also force the \krl system to malfunction by 
modifying the query by adding misleading variables/relations. To make the attack evasive, we require that \mcounter{i} only the addition of variables/relations is allowed and \mcounter{ii} the perturbed query must represent a valid dependency graph with respect to the \kg.

\vspace{2pt}
{\bf Adversary's knowledge --}
%\label{ssec:adversary}
We consider the adversary's knowledge regarding the following two components. 

% \vspace{2pt}
% {\em Knowledge Graph --} Let $\gG$ denote the KG used by the \krl system. As $\gG$ is typically constructed from public sources, we assume the adversary has access to $\gG$.

\vspace{1pt}
{\em Encoder $\phi$ --} The encoder projects \kg entities to their embeddings based on their topological and relational structures. With a little abuse of notation, in the following, we use $\phi_v$, $\phi_\gG$, and   $\phi_q$ to respectively denote the embeddings of entity $v$, \kg $\gG$, and query $q$.

% A plethora of embedding models have been proposed, including vectors\mcite{transe,transh,transr,bilinear-embedding,chains-of-reasoning,kg-traversal,logic-query-embedding}, boxes\mcite{query2box}, and distributions\mcite{beta-embedding}. 

\vspace{1pt}
{\em Operator $\psi$ --} The relation $r$-specific operator $\psi_r$ performs geometric transformation to $\phi_v$ to compute entities with relation $r$ to $v$. For instance, in Query2Box\mcite{query2box}, the \krl consists of two operators, projection and intersection, both implemented as \dnns.  

%reasoning model $\gamma$ takes the query $q$ and the \kg embeddings $\phi_\gG$ as input and processes $q$ over $\gG$: $\gamma(q; \phi_\gG)$. For instance, in the case of entity queries, $\gamma$ consists of two operators, projection and intersection, both implemented as \dnns.  

\begin{table}[!ht]
\renewcommand{\arraystretch}{1.05}
\centering
\setlength{\tabcolsep}{5pt}
{\footnotesize
\begin{tabular}{c|c|c|c|c}
\multirow{3}{*}{Attack} & \multicolumn{2}{c|}{Adversary's Capability} & \multicolumn{2}{c}{ Adversary's Knowledge} \\ 
 \cline{2-5}
  & Knowledge  & Query  & Encoder   & Operator   \\
  & Poisoning  & Perturbation & $\phi$ & $\psi$\\
\hline
\hline
1 & \ding{51} &  &  &  \\
2 & \ding{51} &  & \ding{51}  &  \\
3 & \ding{51} &  &   &  \ding{51}\\
4 & \ding{51} &  & \ding{51}  & \ding{51} \\
\hline
5 &  & \ding{51} &  &  \\
6 &  & \ding{51} & \ding{51}  &  \\
7 &  & \ding{51} &   &  \ding{51}\\
8 &  & \ding{51} & \ding{51}  & \ding{51} \\
\hline
9 & \ding{51} & \ding{51} &  &  \\
10 & \ding{51} & \ding{51} & \ding{51}  &  \\
11 & \ding{51} & \ding{51} &   &  \ding{51}\\
12 & \ding{51} & \ding{51} & \ding{51}  & \ding{51} \\
%\hline
% Attack & knowledge graph (\kg) & \ding{51} & \ding{55} & \ding{51} \\
% vectors & use-case queries & \ding{55} & \ding{51} & \ding{51} \\
% \hline
% \multicolumn{2}{c|}{\multirow{2}{*}{Attack objectives}} & \multicolumn{3}{c}{$\checkmark$ Targeted inference} \\
% \multicolumn{2}{c|}{} & \multicolumn{3}{c}{$\checkmark$ General degradation} \\ 
% \hline
% \multirow{5}{*}{\thead{Adversarial \\ knowledge}} & \ding{192} \kg facts & $\Circle$ & $\LEFTcircle$ & $\Circle$ \\
% & \ding{193} use cases & $\CIRCLE$ & $\Circle$ & $\Circle$\\
% & \ding{194} \kg embedding & $\Circle$ & $\Circle$ &  $\Circle$\\
% & \ding{195} reasoning model & $\LEFTcircle$ & $\LEFTcircle$ & $\LEFTcircle$ \\
% & \ding{196} train queries & $\LEFTcircle$ & $\LEFTcircle$ & $\LEFTcircle$ \\
% \hline
% \multirow{5}{*}{\thead{Adversarial \\ capabilities}} & \ding{192} \kg facts & $\LEFTcircle$ & $\CIRCLE$ & $\LEFTcircle$\\
% & \ding{193} use cases & $\CIRCLE$ & $\LEFTcircle$ & $\LEFTcircle$\\
% & \ding{194} \kg embedding & $\CIRCLE$ & $\CIRCLE$ & $\CIRCLE$\\
% & \ding{195} reasoning model & $\CIRCLE$ & $\CIRCLE$ & $\CIRCLE$\\
% & \ding{196} train queries & $\CIRCLE$ & $\CIRCLE$ & $\CIRCLE$\\
%\hline
\end{tabular}
\caption{A taxonomy of attacks against \krl.\label{tab:taxonomy}}}
\end{table}
\vspace{2pt}

\subsection{Attack taxonomy}
\label{ssec:taxonomy}

% \ting{fix the terms} -- term fixed

Based on the adversary's varying capability and knowledge, we create a taxonomy of 12 attacks against \krl, as summarized in Table\mref{tab:taxonomy}. Given its unique assumption, each attack tends to require a tailored attack strategy. For instance, 

\vspace{1pt}
Attack$_4$ -- It uses knowledge poisoning as the attack vector while assuming knowledge about both encoder $\phi$ and operator $\psi$. Before training, the adversary crafts poisoning facts to pollute the \kg construction in a white-box manner. 

\vspace{1pt}
Attack$_7$ -- It uses query perturbation as the attack vector and assumes knowledge only about $\psi$. Without access to $\phi$, the adversary may resort to a surrogate encoder $\hat{\phi}$ and generate adversarial queries transferable to the target \krl system at inference time.

%At the inference time, the adversary generates adversarial queries to deceive the \krl system. Yet, with access to $\phi$, the adversary may resort to a surrogate model $\phi^*$.  

\vspace{1pt}
Attack$_{10}$ -- It leverages both knowledge poisoning and query perturbation as the attack vectors and assumes knowledge about $\phi$ but not $\psi$; the adversary may thus resort to a surrogate operator $\hat{\psi}$ and optimize the poisoning facts and adversarial queries jointly.

\vspace{1pt}
The discussion of the remaining attacks is deferred to \msec{sec:app:taxonomy}.

\subsection{Challenges}
\label{ssec:challenges}

Note that the attacks against \krl bears conceptual similarity to that against conventional classification models (\meg, adversarial evasion\mcite{goodfellow:fsgm,carlini-attack}, model poisoning\mcite{Suciu:2018:sec,model-reuse}, and backdoor injection\mcite{trojannn,imc}). For instance, both \kg poisoning and model poisoning use poisoning data to influence the training of target systems, while both query perturbation and adversarial evasion aim to generate inputs that are perceptually similar to benign ones but cause target systems to malfunction.

However, realizing the attacks against \krl represents a set of unique, non-trivial challenges. 

\vspace{1pt}
-- Projecting \kg entities or queries (discrete) to their embeddings (continuous) represents a non-differentiable mapping, making it impractical to directly optimize poisoning facts or adversarial queries, which is especially challenging if the encoder is unknown.

\vspace{1pt}
-- There exist potentially combinatorial ways to generate poisoning facts with respect to given \kgs or adversarial queries with respect to given benign queries, which implies a prohibitive search space. 

\vspace{1pt}
-- In attacks that leverage both poisoning facts and adversarial queries, due to their mutual dependence, every time updating the poisoning facts requires expensive re-computation of the adversarial queries. 

%\vspace{2pt}
In the following, we present \system, a new class of attacks against \krl that addresses the above challenges within a novel {\em Optimization-then-Approximation} (\ota) framework.

\section{ROAR Attacks}
\label{sec:attack}

% This section presents technical details of \system attacks, including \kgp, \lqe, and \cop following the \oap framework.

% At the core of \system is {\em optimization-then-projection} (\oap), a novel attack strategy that overcomes the 

Next, we give an overview of the \ota framework and then detail the implementation of KG poisoning and query perturbation within this framework. For simplicity, we assume \mcounter{i} white-box attacks -- the adversary has knowledge regarding the \krl system $\gamma$ (including encoder $\phi$ and operator $\phi$); \mcounter{ii} targeted attacks -- the adversary aims to direct the answering of a query set $\gQ^*$ to the desired answer $A$ (\mie, targeted attacks). We discuss the extension to other settings (\meg, black-box, untargeted attacks) in \msec{sec:extension}. %With a little abuse of notation, we use $\phi_\gG$ and $\phi_q$ to denote the embeddings of $\gG$ and $q$. 

\subsection{Overview of OTA}

Recall that \krl traverses across the input spaces (\mie, KG $\gG$ and query $q$), the latent space (\mie, embeddings $\phi_\gG$ and $\phi_q$), and the output space (\mie, answer $\llbracket q \rrbracket$). To mislead \krl to arrive at the target answer $A$, \ota comprises two key steps: 

\vspace{1pt}
{\em Optimization --} With respect to the desired answer $A$, \ota optimizes the embeddings, $\phi_\gG$ in the case of knowledge poisoning and $\phi_q$ in the case of query perturbation, through backpropagation. Let $\phi_\gG^+$ and $\phi_q^+$ be the updated KG and query.

% To formally describe the {\it optimization-and-projection} (\oap) framework with varying data formats, we first formalize names of three {\it spaces}: the {\it logical space} contains input FOL (\kg or queries); the {\it embedding space} contains latent representations of entities; the {\it output space} refers to returned answers. 

\vspace{1pt}
{\em Approximation --} It then projects $\phi_\gG^+$ (or $\phi_q^+$) back to the input space. However, as $\phi$ is non-differentiable, it leverages attack-specific heuristics to search for KG $\gG^*$ (or query $q^*$) that leads to embeddings to best approximate  $\phi_\gG^+$ (or $\phi_q^+$). 
\vspace{1pt}
These two steps are often executed in an interleaving manner until finding proper poisoning facts (or adversarial queries). Below, we elaborate on the implementation of KG poisoning and query perturbation within the \ota framework.

% \oap traverses from output space back to input space, with two steps to realize attacks: ({\em i}) an {\it optimization} step set up an attack objective (\msec{ssec:objective}), then updates embedding-space objects via backward optimization; ({\em ii}) a {\it projection} step leverages optimized embeddings as supervision, then defines heuristics (based on concrete attacks) to search for corresponding FOL structures. 

% To apply \oap framework on \system (\mie, \kgp, \lqe, or \cop), we need to specify the adversarial operations in each step. Concretely, in the {\it optimization} step, we need to specify the embedding-space objective and to-be-perturbed objects during the backward optimization; in the {\it projection} step, we leverage heuristics (based on specific attacks) to search for logical-space perturbations that best reflects embedding-space results.

% Recap that \system assumes black-box access to a victim system, thus we optimize adversarial strategies on a surrogate system, including embedding-space optimization and logical-space perturbations, then release attacks (\mie, either poisoned facts or query-side perturbations) to influence a victim system. Later, we follow {\it optimization}-to-{\it projection} workflow to elaborate on detailed techniques, where we employ a surrogate system (models and train queries) and use the same notations between the surrogate and the actual ones to ensure clean expressions.

\subsection{Knowledge poisoning}
\label{ssec:kp}

Recall that knowledge poisoning aims to direct the answering of $\gQ^*$ to target answer $A$ by committing poisoning facts $\gF^*$ to the construction of KG $\gG$. 

Without loss of generality, we define $\gQ^*$ as queries sharing a common pattern $p^*$. For instance, $p^*$ can be a specific set of facts forming a logical path
(\meg, $p^* = \textsf{\small Chrome}\xrightarrow[]{\text{vulnerable to}}v_\text{CVE}\xrightarrow[]{\text{fixable by}}v_\text{mitigation}$). In other words, $p^*$ functions as a {\em trigger} to invoke \krl to malfunction. Further, to be practical, we assume the number of poisoning facts that can be added is limited by a threshold $n_\mathrm{kp}$ (\mie, $|\gF^*| \leq n_\mathrm{kp}$).

\begin{figure}[!ht]
    \centering
    \epsfig{file = 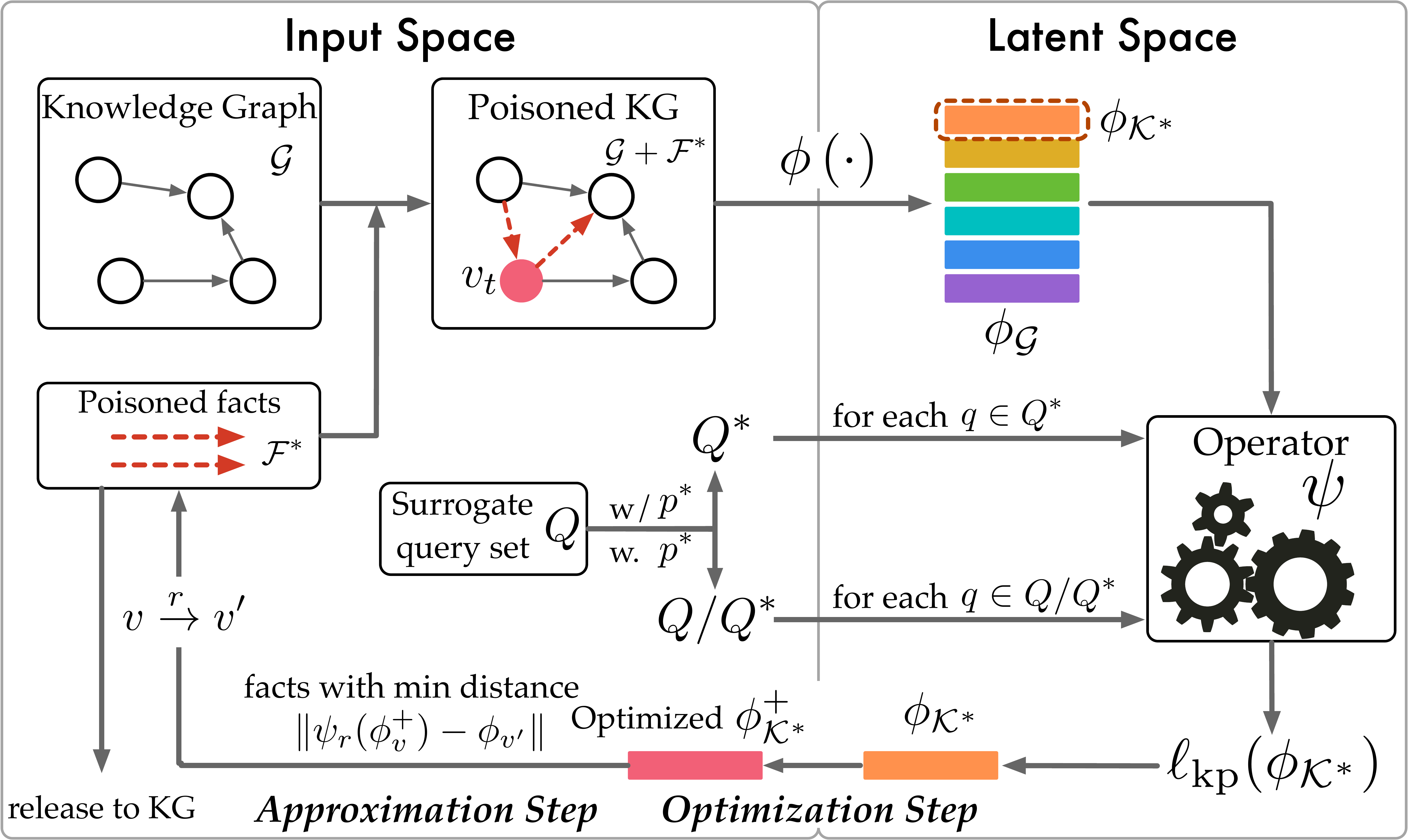, width = 85mm}
    \caption{Workflow of knowledge poisoning.}
    \label{fig:kgp:workflow}
\end{figure}

Let $\gK^*$ collectively denote the anchors entities in $p^*$ (e.g., $\textsf{\small Chrome}$ in the running example) and the target answer $A$. To this end, we define $\gF^*$ as a set of poisoning facts surrounding $\gK^*$ to influence the embeddings of $p^*$, which in turn misleads the reasoning model to malfunction. This overall flow is illustrated in Figure\mref{fig:kgp:workflow}.

\vspace{3pt}
{\bf Optimization --}
%\subsubsection{\textbf{\textit{Optimization}} -- control entity semantics in embedding format}
At this step, we optimize the embeddings $\phi_{\gK^*}$ of $\gK^*$ to achieve the attack objective. Note that adding poisoning facts $\gF^*$ generally influences not only the embeddings of $\gK^*$ but also that of neighboring entities. Yet, due to the locality of  $\gF^*$ surrounding $\gK^*$ and the large scale of $\gG$, we make the assumption that the impact of $\gF^*$ on the embeddings of other entities is marginal, which is also confirmed in our empirical evaluation (\msec{sec:cyber:case}, \msec{sec:drug:case}).

% Perturbing \kg by injecting facts incurs a global influence on embedding semantics, including all entities/relations. Due to the large scale of \kg and multi-hop dependencies, we make an approximation by assuming a local perturbation has ignorable effects on other embeddings, \mie, perturbing local \kg structures of $\ssub{\gV}{tar}$ only falsifies their embeddings, which we denote as $\ssub{\mathbbm{E}}{tar}$.

The optimization of $\phi_{\gK^*}$ carries two objectives: effectiveness -- for query $q \in \gQ^*$, the system returns the desired answer $A$; evasiveness -- for non-target query $q \in \gQ \setminus \gQ^*$, the system returns the ground-truth answer $\llbracket q \rrbracket$. Formally, we define the following loss function: $\ell_\mathrm{kp}(\phi_{\gK^*})$
\begin{equation}
\label{eq:kp}
\hspace{-1pt} = \mathbb{E}_{q \in \gQ^*} \Delta (\gamma(q ; \phi_{\gK^*}),A)
+ \lambda \mathbb{E}_{q \in \gQ \setminus \gQ^*} \Delta (\gamma(q ; \phi_{\gK^*}),\llbracket q \rrbracket)
\end{equation}
where $\Delta$ is a metric (\meg, $L_2$-norm) to measure the distance between two query answers and $\lambda$ is the hyper-parameter to balance the two attack objectives. Note that here we assume the embeddings of all the other entities are fixed.

\vspace{3pt}
{\bf Approximation --} At this step, we project the updated embeddings $\phi_{\gK^*}^+$ back to the input space to search for the optimal poisoning facts $\gF^*$. Recall that the entity embedding function $\phi$ is non-differentiable, it is impractical to directly optimize $\gF^*$. Instead, we adopt a {\em retrograde search} approach to search for $\gF^*$ that leads to embeddings best approximate $\phi_{\gK^*}^+$.

Specifically, we use the relation $r$-specific transformation function $\psi_r$ to assess the quality of candidate facts. Recall that $\psi_r(\phi_v)$ computes entities with relation $r$ to $v$. We enumerate each candidate fact $v \xrightarrow{r} v'$ wherein $v \in \gK^*$ and $v' \in \gN \setminus \gK^*$ under the updated embedding $\phi_v^+$ and select the top-$n_\mathrm{kp}$ candidate facts with the minimum distance of $\| \psi_r(\phi_v^+) - \phi_{v'} \|$. Intuitively, adding such facts to $\gG$ tends to force $\gK^*$ to be projected to $\phi_{\gK^*}^+$. The algorithm is sketched in Algorithm\mref{alg:retro:search}.

% {\it Projection} in \kgp is a process $\ssub{\mathbbm{E}}{tar} \rightarrow \ssub{\gF}{atk}$ such that we use the optimized $\ssub{\mathbbm{E}}{tar}$ from {\it optimization} step as supervision to search for $\ssub{\gF}{atk}$ in logical space. To realize it, we propose a {\it retrograde search} in Algorithm\mref{alg:retro:search} to generate $\ssub{\gF}{atk}$ as logical-space perturbations that best reflect final attack objectives.

\begin{algorithm}[!ht]{\footnotesize
\KwIn{
    $\gN$ -- entities in KG $\gG$; 
    $\gK^*$ -- anchors of trigger pattern $p^*$; 
    $\phi_{v}^+$ -- updated embedding of $v \in \gK^*$;
    $\phi_{v'}$ -- embedding of $v' \in \gN \setminus \gK^*$;
    $\psi_r$ -- relation $r$-specific embedding function;
    $n_\mathrm{kp}$ -- budget of poisoning facts
}
\KwOut{
    $\gF^*$ -- poisoning facts
}
\tcp{initialize}
$\gL \leftarrow \emptyset$\;
\ForEach{$v  \in \gK^*$}{
    \ForEach{$v' \in \gN \setminus \gK^*$}{
        \ForEach{$r \in \gR$}{
        \tcp{compute fitting scores}
        \If{$v \xrightarrow{r} v'$ is legitimate} {
          %\lIf{$|\gF^*| \leq n_\mathrm{poison}}{}
           % score\_map$\left[\left<v, r, \ssub{v}{tar}\right>\right] = \ssub{f}{r}\left(\ssub{\mathbbm{E}}{v}, \ssub{\mathbbm{E}}{t}\right)$ \; 
           compute $\| \psi_r(\phi_v^+) - \phi_{v'} \|$\;
           add $\langle v \xrightarrow{r} v',  \| \psi_r(\phi_v^+) - \phi_{v'} \| \rangle $ to $\gL$\;
        
        %   \lIf{$|\gF^*| \leq n_\mathrm{poison}$}{add $v \xrightarrow{r} v'$ to $\gF^*$}
        %   \ElseIf{$\psi_r(\phi^+_v, \phi_{v'}) \geq \mathrm{score}_\mathrm{min}$}
        %   {
        %     remove the fact with $\mathrm{score}_\mathrm{min}$ from $\gF^*$\;
        %     add $v \xrightarrow{r} v'$ to $\gF^*$\;
        %     update $\mathrm{score}_\mathrm{min}$\;
        %   }
           
        %  or $\psi_r(\phi^+_v, \phi_{v'}) \geq \mathrm{score}_\mathrm{min}$}
        %   {add $v \xrightarrow{r} v'$ to $\gF^*$\;
        %   }
        }
        % \ElseIf{$\left<\ssub{v}{tar}, r, v\right>$ is a legitimate fact} {
        %     score\_map$\left[\left<\ssub{v}{tar}, r, v\right>\right] = \ssub{f}{r}\left(\ssub{\mathbbm{E}}{t}, \ssub{\mathbbm{E}}{v}\right)$\; 
        % }
        }
    }
    % get top $\lfloor \gM / |\ssub{\gV}{tar}| \rfloor$ socres from score\_map \;
    % add corresponding facts to $\ssub{\gF}{atk}$\;
}
sort $\gL$ in descending order of distance \;
\Return the top-$n_\mathrm{kp}$ facts in $\gL$ as $\gF^*$\;
\caption{\small Retrograde Search \label{alg:retro:search}}}
\end{algorithm}

The overall complexity of Algorithm\mref{alg:retro:search} is $O(|\gN||\gK^*||\gR|)$, where $|\gN|$, $|\gK^*|$, and $\gR$ respectively denote the numbers of entities in $\gG$, anchors in trigger pattern $p^*$, and relation types. Further, by taking into account the domain constraints of $\gK^*$ (\meg, certain facts are implausible), the practical complexity tends to be much lower.

\subsection{Query perturbation}
\label{sec:qp}

\begin{figure*}[!ht]
    \centering
    \epsfig{file = 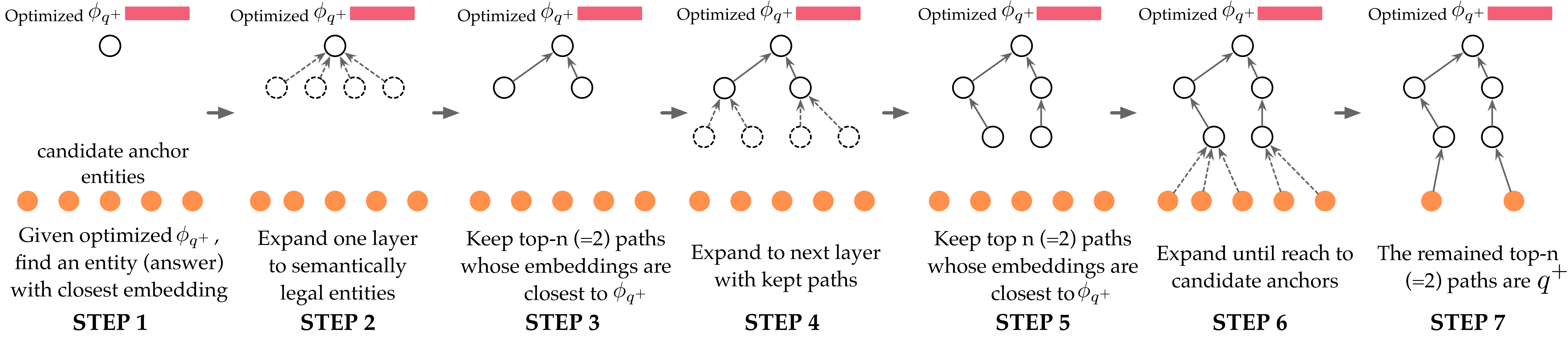, width = 180mm}
    \caption{\small Illustration of constructing $q^+$ via beam search.}
    \label{fig:beam:search}
\end{figure*}

Query perturbation attempts to direct the answering of $\gQ^*$ to target answer $A$ by generating adversarial query $q^*$ from its benign counterpart $q$. We consider $q^*$ as the conjunction of $q$ and additional logical constraint $q^+$: $q^* = q \wedge q^+$. To make the perturbation evasive, we require that \mcounter{i} the number of logical paths in $q^+$ is limited by a threshold $n_\mathrm{qp}$ and \mcounter{ii} the dependency graph of $q^*$ is valid with respect to $\gG$.

Note that while there are potentially combinatorial ways to generate $q^+$, it is often effective to limit  $q^+$ to the vicinity of ground-truth answer $\llbracket q \rrbracket$, which results in more direct influence on altering the query answer. 

At a high level, query perturbation adopts the \ota framework. At the optimization step, we optimize the embedding $\phi_{q^*}$ of $q^*$ with respect to the attack objective; at the approximation step, we further search for potential perturbation $q^+$ that leads to the embedding most similar to $\phi_{q^*}$. Below we elaborate on these two steps.

\vspace{3pt}
{\bf Optimization --} At this step, we optimize the embedding $\phi_{q^*}$ with respect to target answer $A$. Recall that $q^*$ is the conjunction of query $q$ and perturbation $q^+$. We thus define the following loss function: 
\begin{equation}
\label{eq:qp}
\ell_\mathrm{qp}(\phi_{q^+})
 =  \Delta (\psi_{\wedge}(\phi_q, \phi_{q^+}),\, A)
\end{equation}
where $\psi_\wedge$ is the relation $\wedge$-specific transformation function, $\Delta$ is the same distance function as in \meq{eq:kp}, and $\phi_q$ is computed using the current embedding model. We then optimize $\phi_{q^+}$ via backpropagation.

\vspace{3pt}
{\bf Approximation --} At this step, we search for candidate perturbation $q^+$ (input space) that leads to embeddings best approximating $\phi_{q^+}$. 

Recall that in the embedding space $q^+$ is represented as a directed acyclic graph, which starts with the embeddings of its anchor entities, follows the geometric transformation of its relations, and eventually reaches $\phi_{q^+}$. We thus reverse this process to reconstruct $q^+$. However, the complexity of brute-force search (\meg, breadth-first search) grows exponentially with the search depth. Instead, we adopt {\em beam search}\mcite{tillmann2003word}, a greedy strategy, to search for $q^+$.

Specifically, starting with the entity closest to $\phi_{q^+}$ in the latent space as the root, we expand $q^+$ in a level-wise manner until reaching any pre-defined anchor entities. At each level, 
the search enumerates all the neighboring entities in $\gG$, constructs the corresponding  structures, and selects the top-$n_\mathrm{qp}$ candidate structures that lead to embeddings closest to $\phi_{q^+}$. A running example is illustrated in Figure\mref{fig:beam:search}.

The search complexity depends on the depth of $q^+$, the perturbation budget $n_\mathrm{qp}$, and the number of candidate entities in $\gG$. In the worst case, the expansion at each level examines $|\gN|$ candidate structures and selects the top-$n_\mathrm{qp}$ ones, thus featuring the complexity of $O(n_\mathrm{qp} |\gN|)$. As the depth of $q^+$ is less than $\diameter_\gG$, the diameter of $\gG$, the overall complexity is thus $O(n_\mathrm{qp} \diameter_\gG |\gN|)$. In practice, however, as the number of candidate entities is much smaller than $|\gN|$, the practical complexity tends to be much lower.

\subsection{Knowledge-Query co-optimization}
\label{sec:co}

Recall that to direct the answering of a query set $\gQ^*$ (that shares a specific pattern $p^*$) to target answer $A$, the adversary either commits poisoning facts $\gF^*$ to $\gG$ or perturbs the queries in $\gQ^*$. If the adversary is able to leverage both knowledge poisoning and query perturbation, by accounting for the possible perturbation to $\gQ^*$ (at inference time), it is possible to construct more effective poisoning facts $\gF^*$. We thus propose a co-optimization framework that optimizes poisoning knowledge and adversarial queries jointly, leading to more effective attacks.

\begin{algorithm}[!ht]{\footnotesize
\KwIn{
    $\gQ^*$ -- target query set; $\gG$ -- KG; $n_\mathrm{iter}$ -- number of iterations
}
\KwOut{
    poisoning facts to $\gG$; adversarial queries of $\gQ^*$
}
\tcp{initialize}
$\gL \leftarrow \emptyset$\;
$\gQ^{(0)} \leftarrow \gQ^*$\;
\For{$i = 1,\ldots, n_\mathrm{iter}$ }{
\tcp{knowledge poisoning}
generate poisoning facts $\gF^*$ with respect to $\gQ^{(i-1)}$\;
add $\gF^*$ to $\gG$ as $\gG^{(i)}$\;
\tcp{query perturbation}
\ForEach{$q \in \gQ^*$}{
generate adversarial query $q^*$ with respect to $\gG^{(i)}$\;
add $q^*$ to $\gQ^{(i)}$\;
}
}
\Return $\gG^{(n_\mathrm{iter})}\setminus \gG$ and $\gQ^{(n_\mathrm{iter})}$\;
\caption{\small Knowledge-Query Co-optimization \label{alg:co-opt}}}
\end{algorithm}

As sketched in Algorithm\mref{alg:co-opt}, the co-optimization attack iterates between knowledge poisoning and query perturbation: at the $i$-th iteration, in the knowledge poisoning stage, it generates poisoning facts with respect to the current adversarial queries $\gQ^{(i-1)}$; in the query perturbation, it generates adversarial queries with respect to the updated KG $\gG^{(i)}$. The process iterates until convergence.

\subsection{Extension}
\label{sec:extension}

We now discuss the extension of \system to other settings. 

\vspace{1pt}
{\em Black-box attacks --} Without knowledge regarding encoder $\phi$ or operator $\psi$, the adversary may resort to surrogate models to approximate $\phi$ or $\psi$. For instance, it is empirically shown that TransE\mcite{transe} and TransR\mcite{transr}, two widely used entity embedding models, have fairly similar behavior. It is thus possible to generate poisoning facts/adversarial queries on the surrogate model and then transfer the attack to the target system. 

\vspace{1pt}
{\em Untargeted attacks --} An untargeted attack aims to direct the answering of target queries $\gQ^*$ to erroneous answers rather than a specific answer $A$. Thus, at the optimization stage of the attack (\meq{eq:kp} and \meq{eq:qp}), instead of minimizing the distance between the answer of $q \in \gQ^*$ and $A$, the adversary may maximize the distance between the answer of $q$ and its ground-truth answer $\llbracket q \rrbracket$. For instance, we may re-define \meq{eq:qp} as: 
\begin{equation}
\label{eq:untargetd}
\ell_\mathrm{qp}(\phi_{q^+})
 =  -\Delta (\psi_{\wedge}(\phi_q, \phi_{q^+}),\, \llbracket q \rrbracket)    
\end{equation}

\section{Case Study I: Cyber-Threat Hunting}
\label{sec:cyber:case}

Next, we conduct an empirical evaluation of \system in the concrete case of cyber-threat intelligence reasoning. Our study is designed to answer the following questions: Q$_1$ -- Is \system effective against \krl systems in practice?
Q$_2$ -- How does the effectiveness of its variants differ? Q$_3$ -- What factors impact the performance of \system?

\subsection{Experimental setup}

We begin by describing the evaluation setting.

\vspace{2pt}
{\bf Scenario --} With the explosive growth of cyber-threat intelligence, it becomes imperative for security analysts to rely on automated tools to extract useful, actionable intelligence for given incidents\mcite{mittal2016cybertwitter, mittal2019cyber}. Here, we consider a \krl system built upon cyber-threat KGs (\meg, attack tactics, vulnerabilities, fixes), which supports the following queries: 

\vspace{1pt}
{\em Vulnerability --} Given certain observations regarding the incident (\meg, affected products/versions, attack tactics, or campaigns), it finds the most likely vulnerabilities (\meg, \cves) that are being exploited.

\vspace{1pt}
{\em Mitigation --} Beyond the vulnerabilities being attacks, it may further suggest potential solutions (\meg, patches and workarounds) to mitigate such vulnerabilities. 

\vspace{1pt}
Besides querying for known vulnerabilities, we also consider a zero-day setting wherein the attacks exploit previously unknown vulnerabilities.

\vspace{2pt}
{\bf CyberKG --} We construct the CyberKG using cyber-threat intelligence from three sources: ({\em i}) public \cve reports\footnote{https://www.cvedetails.com}; ({\em ii}) BRON threat graph\mcite{bron}; ({\em iii}) National Vulnerability Database ({\small NVD})\footnote{https://nvd.nist.gov}.

\vspace{1pt}
-- From ({\em i}), we collect known {\sl CVE} as potential vulnerabilities, each associated with {\sl affected product}, {\sl version}, {\sl vendor}, {\sl CWE}\footnote{https://cwe.mitre.org}, and {\sl threat campaign}. 

\vspace{1pt}
-- From ({\em ii}), we extract {\sl attack tactic}, {\sl technique} and {\sl pattern}, and link them with ({\it i}) based on \cve codes.

% the BRON graph based on common CVEs in ({\it i}), in which we keep entities related to tactics, techniques, attack patterns, and weaknesses.

\vspace{1pt}
-- From ({\it iii}), we collect {\sl mitigation} for each \cve and assign a specific code to each mitigation approach. % Then we attach mitigation codes to the constructed \kg based on their corresponding CVEs.

\vspace{1pt}
The logical structures of the resulted \kg is illustrated in Figure\mref{fig:cyberkg}, with its statistics summarized in Table\mref{tab:cyber:stat}.

\vspace{2pt}
{\bf Query -- } To construct queries, for given {\sl CVE} or {\sl mitigation}, we randomly select 
anchors from the remaining entities (\meg, {\sl affected product}, {\sl attack pattern}, and {\sl threat campaign}) and formulate the queries based on their logical structures in the KG. 
Figure\mref{fig:cyber:qpath} shows the logical paths with different anchor entities, which are combined to formulate the query templates as shown in Figure\mref{fig:cyber:qstruc}. In the evaluation,  we generate 500 queries for each query template.

% consider the collectible evidence to construct queries. For example, it is easy to get product names, versions, and belonged vendors by checking product instructions. Moreover, we can identify the adversarial techniques, attack patterns, and campaigns through (software) testing. We then use those evidence as anchor entities and construct logical paths based on our \kg structure. 

% As a summary, we have 500 queries for each test query structure (4500 test queries in total) and have 160000 train queries in total. More details are presented in Appendix \msec{ssec:qstruc}

\vspace{2pt}
{\bf KRL --} We apply Query2Box\mcite{query2box}, a state-of-the-art \krl approach, to build the reasoning system. Intuitively, by adopting box embeddings (\mie, hyper-rectangles), Query2Box naturally handles first-order conjunctive queries. 

% The threat intelligence case leverages a state-of-the-art box embedding named  for both \kg embedding and reasoning. During reasoning, Query2Box processes each query by its directed acyclic structure, transfers it into box embedding format, then matches ontological answers in embedding space. This paradigm fits the threat intelligence case since it is an ontological reasoning problem.

%\vspace{2pt}
\begin{table}[!ht]
\renewcommand{\arraystretch}{1.2}
\centering
{\small
 \begin{tabular}{c|cc}
%  \toprule
% \Xhline{1.5\arrayrulewidth}
 Query &  MRR & NDCG@5 \\ 
\hline 
\hline
Vulnerability & 0.61 &  0.76 \\ 
Mitigation & 0.47 &   0.66 \\
 \end{tabular}
 
\caption{\krl performance in cyber-threat hunting. \label{tab:benign:all}}}
\end{table}

\vspace{2pt}
{\bf Metrics --} In this case study, we mainly use two metrics to evaluate the performance of \krl and attacks:
% threat intelligence is an ontological reasoning task that contains more than 18k threat codes are and more than 74k mitigation entities, hence we use $MRR$ and $NDCG@5$ to show rank qualities of all ground-truth answers and top-5 results, respectively. 

{\mrr} -- It computes the average reciprocal ranks of all the ground-truth answers across all the queries. This metric measures the global ranking quality of reasoning answers. 

{\ndcg} -- It evaluates the ranking quality of the top-$K$ answers using the NDCG metric\mcite{metrics}.

All the measurements range from 0 to 1, with larger values indicating better performance. Table\mref{tab:benign:all} summarizes the performance of the \krl system in cyber-threat hunting.

% ({\em i}) \textbf{\textit{MRR}} evaluates the mean reciprocal ranks of all answers among all queries. This metric concerns a global rank quality of reasoning answers. 
% ({\em ii}) \textbf{\textit{HIT@K}} evaluates whether top-$K$ results contain ground-truth answers. For each query, this metric gives a binary indication (0/1). We average those binary results and get a global hits ratio.
% ({\em iii}) \textbf{\textit{NDCG@K}} evaluates the ranking quality of top-$K$ results with the NDCG metric\mcite{}\ting{cite sth}, which is more strict than {\it HIT@K} since positions of ground-truth answers in top-$K$ ranks affects its value.

\subsection{Attack implementation}

\label{sec:case1:attack}

Next, we detail the implementation of \system attacks in the case study of cyber-threat hunting.

\vspace{2pt}
{\bf Attack settings --} By default, we set target queries $\gQ^*$ as that containing a specific pattern $p^* = \textsf{\small Android OS} \xrightarrow[]{\text{vulnerable to}} v_\text{CVE} \xrightarrow[]{\text{fixable by}} v_\text{mitigation}$, where 
$\textsf{\small Android OS}$ is an anchor entity while both $v_\text{CVE}$ and $v_\text{mitigation}$ are variables. In \msec{sec:case1:eval}, we also consider alternative definitions of $p^*$.

We consider both targeted and untargeted attacks. In untargeted attacks, the goal is to direct the answering of each query $q \in \gQ^*$ to an erroneous one (different from its ground-truth $\llbracket q \rrbracket$); in targeted attacks, the goal is to direct the answering of all the queries in $\gQ^*$ to a target answer $A$. For vulnerability queries, we set $A = \textsf{\small \cve-2021-0471}$, which is a \cve with lower severity among all the Android \cves; for mitigation queries, we set $A$ as the mitigation of $\textsf{\small CVE-2021-0471}$.

Further, it is necessary to ensure the impact of the attacks on benign queries (\mie, without $p^*$) is limited.

% \subsubsection{Attack objectives} 

% In this case, we falsify reasoning results among queries with targeted logical path $\ssub{\gL}{tar}$, which is $\mathsf{Android\,\, OS}\xrightarrow[]{\text{suffers from}}\left<\mathsf{CVE?}\right>\xrightarrow[]{\text{be fixed by}}\left<\mathsf{Mitigation?}\right>$ by default. Queries with the anchor entity $\mathsf{Android\,\, OS}$ and the following  path $\xrightarrow[]{\text{suffers from}}\left<\mathsf{CVE?}\right>\xrightarrow[]{\text{be fixed by}}\left<\mathsf{Mitigation?}\right>$ are $\ssub{Q}{tar}$, and will mislead the threat engine into erroneous results, which include incorrect threat codes (\mie, CVEs) and mitigation approaches. We denote those erroneous answers as $\ssub{A}{atk}$.

% $\ssub{A}{atk}$ has different specifications in different attack objectives: ({\em i}) in targeted inference attacks, we specify $\ssub{A}{atk}$ as CVE-2021-0471 when querying threat code.  CVE-2021-0471 refers to a less vulnerable threat among all Android CVEs. We specify corresponding mitigation approaches as targeted answers when querying mitigation. ({\em ii}) in general degradation attacks, we have no specific expectation as long as reasoning results depart from ground-truth answers (CVEs or mitigation).

% Among queries $\ssub{Q}{\backslash tar}$ without $\ssub{\gL}{tar}$, attacks with both objectives should not affect the reasoning performances, thus ensuring the victim threat engine is still credible in most use cases.

% \subsubsection{Attack configurations}

{\bf Attack variants --} To the best of our knowledge, this is the first work on the security of \krl. We thus mainly compare the performance of different variants of \system:

\vspace{1pt}
\akp -- Relying on knowledge poisoning, the adversary influences \krl by committing poisoning facts to the \kg construction, but has no control over the queries. Specifically, \akp commits poisoning facts surrounding $\textsf{\small Android OS}$ (anchor entity of $p^*$) and $\textsf{\small CVE-2021-0471}$ (target answer $A$). We limit the number of poisoning facts by $n_\mathrm{kp} = 100$.

% The \underline{s}tand\underline{a}lone \kgp attack. The attacker is only accessible to \kg as a malicious contributor. A poisoned \kg affects downstream embedding and reasoning without further manipulation of the \kgp attacker.

\vspace{1pt}
\aqp -- Relying on query perturbation, the adversary generates adversarial queries at inference time but has no control over the \kg construction. To make the perturbation evasive, we require that \mcounter{i} the number of logical paths in the perturbation is limited by  $n_\mathrm{qp} = 2$ and \mcounter{ii} the dependency graph of the perturbed query is valid with respect to the \kg.

%The \underline{s}tand\underline{a}lone \lqe attack. The attacker locates at downstream use cases, thus only affects query contents, including anchor entities and logical structures.

\vspace{1pt}
\aco  -- Leveraging both knowledge poisoning and query perturbation, the adversary optimizes the poisoning facts and adversarial queries jointly. 

% The \underline{c}o-\underline{o}ptimization attack between \\kgp and \lqe, where the downstream \lqe attacker returns specific anchor entities to \kgp attacker for stronger poisoning. Both \kgp and \lqe co-optimize their attacks on a substitute system.

\vspace{1pt}
In all the attacks, we assume the adversary has no control over the training of \krl including \mcounter{i} how the training set is sampled from the \kg and \mcounter{ii} how the entity embedding and relation transformation models are trained. Thus, all the poisoning facts and/or adversarial queries are crafted on surrogate models.

% The \\kgp attacker reports malicious facts to influence local \kg structures 
% related to specific entities. All reported facts contain legal relations; hence they can pass verifications and immigrate into \kg. The standalone \kgp attack $\ssub{SA}{kgp}$ does not know downstream queries, thus only perturbs the anchor entity in $\ssub{\gL}{tar}$, \mie $\mathsf{Android\,\,OS}$, and its 1-hop neighbor entities in \kg. In targeted inference,  \kgp also adds facts surrounding the targeted threat code (CVE-2021-0471) or corresponding mitigation entities. We constrain the number of poisoning facts within a budget $\gM=100$ by default. 

% \lqe happens when a security analyst tests products (\meg, software applications) to collect evidence. In reality, the attacker injects malicious functions beforehand, which cause the product to perform additional behaviors, then collected by the analyst as part of query structures. In experiments, the \lqe perturbation only leverages entities (and their following logical paths) that belong to the same product of a victim query, such as different versions, or adds technique-related entities that simulate intentional threat behaviors. By default, the evasion scale $\gS=2$, \mie, we inject two additional logical paths as perturbations.

% The \cop attack is capable of both \kg poisoning and downstream evasion. Hence we iteratively make \kgp and \lqe attacks on a substitute system until attack performances remain stable. Then we add the searched facts to poison \kg, together with downstream evasion perturbations, and evaluate final results.

\subsection{Evaluation results}
\label{sec:case1:eval}

For exposition simplicity, below we highlight the \krl performance variation before and after the attacks. In targeted attacks, ``$\uparrow$'' indicates the score increase (\meg, \mrr) with respect to target answer $A$; in untargeted attacks, ``$\downarrow$'' indicates the score decrease with respect to ground-truth answers.

% In this part, we evaluate attack performances on two applications -- querying threat codes and querying mitigation -- with both targeted inference and general degradation objectives. Each query in training/evaluation set has two groups of answers, \mie, CVEs and mitigation entities. When querying mitigation, the reasoning system treats CVEs as variable entities, instead of using CVEs to search for mitigation results.

% In later evaluations, we show performance variations before and after attacks under different objectives, which are: ({\it i}) increasing values followed with ``$\uparrow$'' during targeted inference; ({\it ii}) decreasing values followed with ``$\downarrow$'' during general degradation.

\begin{table}[!ht]
\renewcommand{\arraystretch}{1.2}
\centering
\setlength{\tabcolsep}{2pt}
{\footnotesize
\begin{tabular}{ c|c|lr|lr|lr|lr}
 \multirow{2}{*}{Objetive} & \multirow{2}{*}{Query} & \multicolumn{8}{c}{Attack} \\ 
 \cline{3-10} 
  & & \multicolumn{2}{c|}{w/o} &\multicolumn{2}{c|}{\akp} & \multicolumn{2}{c|}{\aqp} & \multicolumn{2}{c}{\aco} \\
  \hline
  \hline
 \multirow{2}{*}{Targeted} & Vulnerability & .00 & .00 & .53$\uparrow$ & .57$\uparrow$ & \multicolumn{1}{c}{\NA} & \multicolumn{1}{c|}{\NA} & .99$\uparrow$ & .99$\uparrow$ \\
   &  Mitigation & .00 & .00 & .29$\uparrow$ & .33$\uparrow$ & \multicolumn{1}{c}{\NA} & \multicolumn{1}{c|}{\NA} & .58$\uparrow$ & .60$\uparrow$ \\
  \hline
 \multirow{2}{*}{Untargeted} & Vulnerability & .56 & .75 & .34$\downarrow$ & .22$\downarrow$ & .49$\downarrow$ & .48$\downarrow$ & .55$\downarrow$ & .73$\downarrow$\\
  &  Mitigation & .48 & .68 & .31$\downarrow$ & .20$\downarrow$ & .45$\downarrow$ & .57$\downarrow$ & .48$\downarrow$ & .68$\downarrow$ \\
 \hline
 \end{tabular}
\caption{Overall attack performance in cyber-threat hunting. Note: the values in each cell are MRR (left) and NDCG@5 (right).\label{tab:cyber:attack:overall}}}
\end{table}

\vspace{2pt}
{\bf Attack effectiveness --} Table\mref{tab:cyber:attack:overall} summarizes the overall attack performance measured by \mrr and \ndcgf. Note that without attacks (w/o), none of the queries in $\gQ^*$ leads to target answer $A$, due to their irrelevance. We have the following interesting observations.

% In the beginning, we make no attack on the reasoning system, hence the targeted inference on $\ssub{Q}{tar}$ has no performance because the targeted answer is totally irrelevant with all queries in  $\ssub{Q}{tar}$; and reasoning performances on $\ssub{Q}{tar}$ (with ground-truth answers) are also normal comparing with overall benign results in Table\mref{tab:benign:all}. 

% When perturbing \kg ($SA_{kgp}$), downstream queries ($SA_{lqe}$), or both $CoP$, attacks benign to be effective. We use $\uparrow$ in targeted inference to show the increased attack performance and $\downarrow$ in general degradation to show the decreased value. Both compare with the no-attack scenario. Here are some observations from Table\mref{tab:attack:cyber}:

\underline{\textit{\akp is more effective in targeted attacks.}} -- \aqp seems unable to direct \krl to target answer $A$. This may be explained by that limited by the perturbation constraints (\msec{sec:qp}), it is difficult to direct the answering of query $q$ to $A$ remotely relevant to $\llbracket q \rrbracket$. In contrast, \akp exerts a more significant influence on \krl by minimizing the distance between the embeddings of pattern $p^*$ (shared by $\gQ^*$) and $A$, leading to more effective targeted attacks.

% It injects a limited number of evasion logical paths to a query, hence cannot induce a benign model to a targeted answer irrelevant to the original query. While $SA_{kgp}$ is more effective than $SA_{lqe}$ because it affects the model by poisoning the embedding space. $SA_{kgp}$ controls embedding semantics of some entities (\meg, $\mathsf{Android\,\,OS}$ from $\ssub{\gL}{tar}$) by perturbing their local \kg structure (\mie, FOL), thus injects bias to the reasoning model when training with those embeddings, which further makes $\ssub{\gL}{tar}$ more influential than other logical paths in a query and misleads the poisoned model to a targeted answer.

\underline{\textit{\aqp is more effective in untargeted attacks.}} -- Interestingly, \aqp outperforms \akp in untargeted attacks. This may be explained as follows. An untargeted attack succeeds if the answer of query $q$ deviates from $\llbracket q \rrbracket$; the poisoning facts of \akp affects all the queries in the same manner, while in \aqp, the perturbation is tailored to each query $q$, resulting in more effective untargeted attacks.

% the general degradation succeeds as long as reasoning answers are different from ground-truth ones, thus easier to realize comparing with the targeted inference. The $SA_{lqe}$ is effective on a benign model since it searches and injects logical paths to a victim query so that the reasoning model cannot correctly identify the query in embedding format. However, even though $SA_{kgp}$ falsifies embedding semantics of some anchor entities in a query, the reasoning model can still infer answers based on other logical paths, hence $SA_{kgp}$ is less effective than $SA_{lqe}$.

\underline{\textit{\aco is the most effective attack.}} -- In both targeted and untargeted cases, \aco outperforms the other attacks. For instance, in targeted attacks against vulnerability queries, \aco attains 0.99 increase in \mrr. This may be attributed to the mutual reinforcement between knowledge poisoning and query perturbation: tailoring poisoning facts to adversarial queries, and vice versa, improves the attack effectiveness.

% \ting{consider removing this paragraph}

% \underline{\textit{Attacking the longer path is harder}} -- table \mref{tab:benign:all} and \mref{tab:attack:cyber} indicate that, both benign reasoning and attacks are harder on querying mitigation because the logical structure used for querying mitigation is 1-hop longer than the one for querying \cves. In embedding-space calculations, the longer relational structure introduces more noises hence causes query embeddings more inaccurate, which brings more difficulties to match correct reasoning answers, also keeps less poisoning effects in embedding format.

\begin{table}[!ht]
\renewcommand{\arraystretch}{1.2}
\centering
\setlength{\tabcolsep}{4pt}
{\footnotesize
\begin{tabular}{c|c|lr|lr}
 \multirow{2}{*}{Objective} & \multirow{2}{*}{Query} & \multicolumn{4}{c}{Attack} \\ 
 \cline{3-6} 
%  \cmidrule(lr){5-8}
  & & \multicolumn{2}{c|}{\akp} & \multicolumn{2}{c}{\aco} \\
  \hline
  \hline
  \multirow{2}{*}{Targeted} & Vulnerability & 0.00$\downarrow$  & 0.00$\downarrow$ & 0.00$\downarrow$ & 0.00$\downarrow$ \\
   & Mitigation & 0.01$\downarrow$ & 0.00$\downarrow$ & 0.00$\downarrow$ & 0.02$\downarrow$ \\
  \hline
 \multirow{2}{*}{Untargeted} & Vulnerability & 0.00$\downarrow$ & 0.02$\downarrow$ & 0.00$\downarrow$  &  0.00$\downarrow$ \\
 & Mitigation  & 0.01$\downarrow$  & 0.01$\downarrow$ & 0.00$\downarrow$ & 0.00$\downarrow$ \\
 \hline
 \end{tabular}
\caption{\small Impact of attacks on non-target queries. Note: the values in each cell are MRR (left) and NDCG@5 (right). \label{tab:benign:drop:cyber}}}
\end{table}

\vspace{2pt}
{\bf Attack evasiveness --} We further measure the impact of the attacks on non-target queries (without pattern $p^*$). As \aqp has no influence on non-target queries, we focus on evaluating \akp and \aco, with results shown in  Table\mref{tab:benign:drop:cyber}. 

\underline{{\em \system has a limited impact on non-target queries.}} -- Compared with the benign system ({\em cf.} Table\mref{tab:benign:all}), \akp and \aco have negligible influence on the answering of non-target queries. This may be attributed to multiple factors: \mcounter{i} the limited number of poisoning facts (less than $n_\mathrm{kp}$), \mcounter{ii} the locality of such facts, and 
\mcounter{iii} the large size of \kg.

% Model-poisoning attacks should not influence the normal reasoning ability w.r.t. queries without $\ssub{\gL}{tar}$, thus ensure evasiveness and not make the reasoning system useless in practical. In Table\mref{tab:benign:drop:cyber}, we present benign reasoning performances on $\ssub{Q}{\backslash tar}$ that does not contain  $\ssub{\gL}{tar}$. Comparing with original performances , the {\it MRR} decreases 0.01 and {\it NDCG@5} decreases 0.02 at most; hence we demonstrate that attacks ($SA_{kgp}$, $CoP$) has ignorable effects to the benign reasoning ability of poisoned models.

\begin{table*}[!ht]
\renewcommand{\arraystretch}{1.2}
\centering
\setlength{\tabcolsep}{2pt}
{\footnotesize
\begin{tabular}{ cccccc|cc|cccccc|cc|cccccc|cc|cc}
%  \toprule
\hline
 \multirow{2}{*}{Objective} &  \multirow{2}{*}{Attack} & & & \multicolumn{22}{c}{ Query Topological Structure ($n_\mathrm{path}$-$m_\mathrm{path}$)} \\ 
 \cline{5-26} 
%  \cmidrule(lr){5-26}
  & & & &
  \multicolumn{2}{c}{3-1} & \multicolumn{2}{c}{3-2} & \multicolumn{2}{c}{3-3} & 
  & &
  \multicolumn{2}{c}{5-1} & \multicolumn{2}{c}{5-2} & \multicolumn{2}{c}{5-3} &
  & &
  \multicolumn{2}{c}{7-1} & \multicolumn{2}{c}{7-2} & \multicolumn{2}{c}{7-3} \\
%   \midrule
\hline
\hline
 & w/o & & & 0.00 & 0.00 & 0.00 & 0.00 & 0.00 & 0.00 & & & 0.00 & 0.00 & 0.00 & 0.00 & 0.00 & 0.00 & & & 0.00 & 0.00 & 0.00 & 0.00 & 0.00 & 0.00 \\
\multirow{2}{*}{Targeted} & \akp & & & 0.51$\uparrow$ & 0.56$\uparrow$ & 0.78$\uparrow$ & 0.81$\uparrow$ & 0.78$\uparrow$ & 0.81$\uparrow$ & & & 0.37$\uparrow$ & 0.40$\uparrow$ & 0.63$\uparrow$ & 0.71$\uparrow$ & 0.66$\uparrow$ & 0.71$\uparrow$ & & & 0.23$\uparrow$ & 0.27$\uparrow$ & 0.38$\uparrow$ & 0.42$\uparrow$ & 0.39$\uparrow$ & 0.44$\uparrow$ \\
& \aqp & & & \multicolumn{1}{c}{\NA} & \multicolumn{1}{c|}{\NA} & \multicolumn{1}{c}{\NA} & \multicolumn{1}{c|}{\NA} & \multicolumn{1}{c}{\NA} & \multicolumn{1}{c}{\NA} & & & \multicolumn{1}{c}{\NA} & \multicolumn{1}{c|}{\NA} & \multicolumn{1}{c}{\NA} & \multicolumn{1}{c|}{\NA} & \multicolumn{1}{c}{\NA} & \multicolumn{1}{c}{\NA} & & & \multicolumn{1}{c}{\NA} & \multicolumn{1}{c|}{\NA} & \multicolumn{1}{c}{\NA} & \multicolumn{1}{c|}{\NA} & \multicolumn{1}{c}{\NA} & \multicolumn{1}{c}{\NA} \\
& \aco & & & 0.98$\uparrow$ & 0.98$\uparrow$ & 1.00$\uparrow$ & 1.00$\uparrow$ & 1.00$\uparrow$ & 1.00$\uparrow$ & & & 0.97$\uparrow$ & 0.98$\uparrow$ & 1.00$\uparrow$ & 0.99$\uparrow$ & 1.00$\uparrow$ & 1.00$\uparrow$ & & & 0.94$\uparrow$ & 0.96$\uparrow$ & 1.00$\uparrow$ & 1.00$\uparrow$ & 1.00$\uparrow$ & 1.00$\uparrow$ \\
% \midrule
\hline
 & w/o & & & 0.59 & 0.83 & 0.52 & 0.75 & 0.52 & 0.74 & & & 0.60 & 0.78 & 0.56 & 0.75 & 0.54 & 0.72 & & & 0.60 & 0.76 & 0.56 & 0.72 & 0.58 & 0.72 \\
\multirow{2}{*}{Untargeted} & \akp & & & 0.35$\downarrow$ & 0.24$\downarrow$ & 0.31$\downarrow$ & 0.30$\downarrow$ & 0.42$\downarrow$ & 0.40$\downarrow$ & & & 0.34$\downarrow$ & 0.19$\downarrow$ & 0.34$\downarrow$ & 0.25$\downarrow$ & 0.35$\downarrow$ & 0.28$\downarrow$ & & & 0.31$\downarrow$ & 0.12$\downarrow$ & 0.33$\downarrow$ & 0.09$\downarrow$ & 0.36$\downarrow$ & 0.11$\downarrow$ \\
& \aqp & & & 0.50$\downarrow$ & 0.54$\downarrow$ & 0.51$\downarrow$ & 0.66$\downarrow$ & 0.51$\downarrow$ & 0.65$\downarrow$ & & & 0.50$\downarrow$ & 0.41$\downarrow$ & 0.49$\downarrow$ & 0.46$\downarrow$ & 0.47$\downarrow$ & 0.44$\downarrow$ & & & 0.50$\downarrow$ & 0.37$\downarrow$ & 0.47$\downarrow$ & 0.41$\downarrow$ & 0.49$\downarrow$ & 0.41$\downarrow$ \\
& \aco & & & 0.58$\downarrow$ & 0.81$\downarrow$ & 0.52$\downarrow$ & 0.75$\downarrow$ & 0.52$\downarrow$ & 0.74$\downarrow$ & & & 0.58$\downarrow$ & 0.74$\downarrow$ & 0.54$\downarrow$ & 0.73$\downarrow$ & 0.53$\downarrow$ & 0.71$\downarrow$ & & & 0.57$\downarrow$ & 0.70$\downarrow$ & 0.55$\downarrow$ & 0.71$\downarrow$ & 0.57$\downarrow$ & 0.71$\downarrow$ \\
 \hline
 \end{tabular}
\caption{\small Topological complexity of queries on the attack performance. Note: the values in each cell are MRR (left) and NDCG@5 (right). \label{tab:cyber:attack:diff:struc}}}
\end{table*}

\vspace{2pt}
\textbf{Query structures --} Recall that the queries are generated following a set of templates ({\em cf.} Figure\mref{fig:cyber:qstruc}). We now evaluate the impact of query structures on the attack performance. We encode the complexity of query $q$ as $n_\mathrm{path}$-$m_\mathrm{path}$, where 
$n_\mathrm{path}$ is the number of logical paths (from anchors $\gK_q$ to answer $\llbracket q \rrbracket$) in $q$ and $m_\mathrm{path}$ is the length of the longest path. Table\mref{tab:cyber:attack:diff:struc} breaks down the attack performance according to the setting of $n_\mathrm{path}$ and $m_\mathrm{path}$. We have the following observations.

% also concern how attacks perform with different query structures . From evaluation results in Table\mref{tab:cyber:attack:diff:struc}, we observe that the number of logical paths and complexity level affects attack efficacy with specific trends: 

\underline{{\em Attack performance drops with $n_\mathrm{path}$.}} -- By increasing the number of logical paths $n_\mathrm{path}$ but keeping the maximum path length $m_\mathrm{path}$ fixed, the effectiveness of all the attacks tends to drop. This may be explained as follows. Each logical path in query $q$ represents one constraint on the answer $\llbracket q \rrbracket$; with more constraints, \krl is more robust to local perturbation to either \kg or parts of $q$.

% ({\em i}) When increasing logical path numbers (3$\rightarrow$5$\rightarrow$7) while keeping the same complexity level (I/II/III), all attacks lose their efficacies due to more evidence present in queries, hence the reasoning is more robust even though some paths are perturbed by attackers.

\underline{{\em Attack performance improves with $m_\mathrm{path}$ in targeted cases.}} -- Interestingly, in the targeted cases, the attack performance improves with $m_\mathrm{path}$ under fixed $n_\mathrm{path}$. This may be explained as follows. Longer logical paths in $q$ represent ``weaker'' constraints, due to the accumulated approximation errors of relation transformation. As $p^*$ is defined as a short logical path, for queries with large $m_\mathrm{path}$, $p^*$ tends to dominate the query answering, resulting in more effective attacks.

% ({\em ii}) When increasing complexity levels (I$\rightarrow$II$\rightarrow$III) and keep with a same logical path number (3/5/7), all attacks become easier to succeed especially in targeted inference. The reasoning model relies more on the shorter logical paths since longer paths cannot lead to accurate embedding representations. Our $\ssub{\gL}{tar}$ is a short path (1-hop to \cves and 2-hop to mitigation entities); hence the reasoning model relies more on it, which leads to a more effective attack when queries are more complex (\mie, with more multi-hop paths).

\begin{figure*}
    \centering
    \epsfig{file = 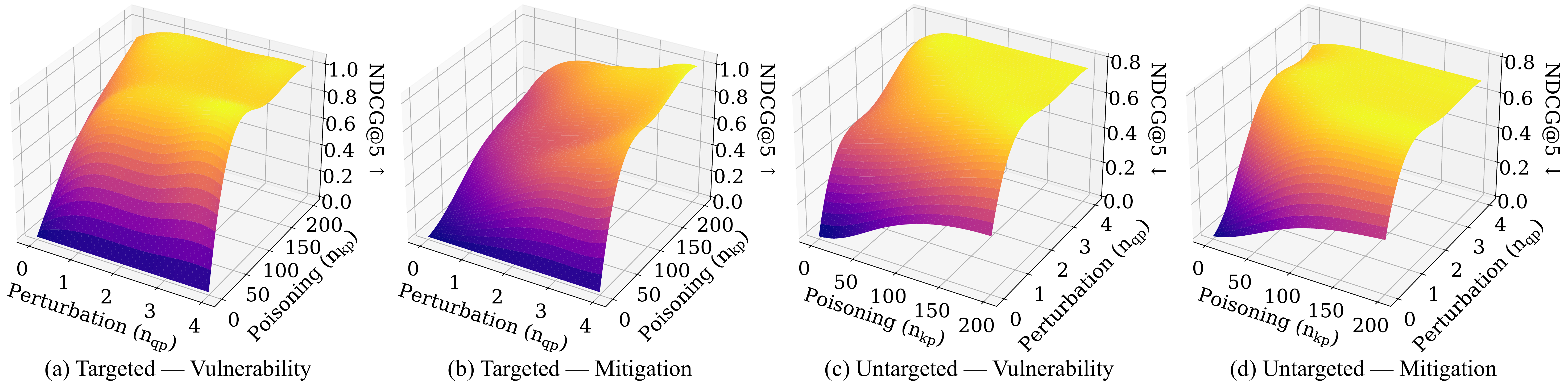, width = 170mm}
    \caption{NDCG@5 variation of \aco as a function of knowledge poisoning budget ($n_\mathrm{kp}$) and query perturbation budget ($n_\mathrm{qp}$).}
    \label{fig:cyber:cop:ndcg}
\end{figure*}

\begin{figure*}
    \centering
    \epsfig{file = 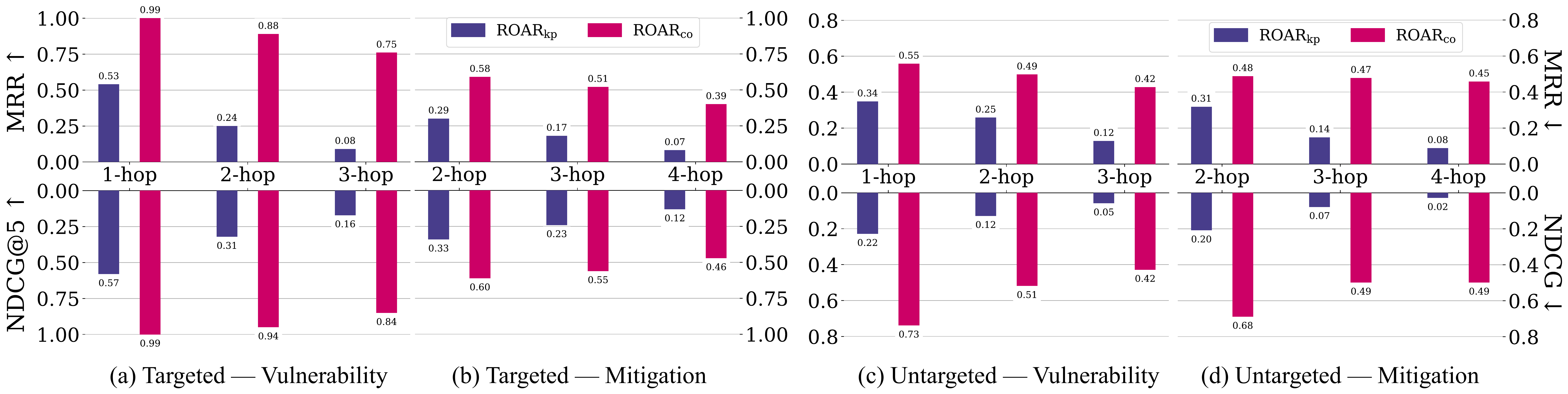, width = 170mm}
    \caption{Attack performance under alternative definitions of $p^*$.}
    \label{fig:cyber:diff:tar:logic}
\end{figure*}

\begin{table}[!ht]
\renewcommand{\arraystretch}{1.2}
\centering
\setlength{\tabcolsep}{2.5pt}
{\footnotesize
\begin{tabular}{ c|c|lr|lr|lr}
 \multirow{2}{*}{Objective} & \multirow{2}{*}{Query} & \multicolumn{6}{c}{Attack} \\ 
 \cline{3-8} 
  & &\multicolumn{2}{c|}{\akp} & \multicolumn{2}{c|}{\aqp} & \multicolumn{2}{c}{\aco} \\
  \hline
  \hline
  \multicolumn{8}{c}{Embedding Dimensionality = 200, DNN Depth = 1} \\
   \hline
 \multirow{2}{*}{Targeted} & Vulnerability & .48$\uparrow$ & .55$\uparrow$ & \multicolumn{1}{c}{\NA} & \multicolumn{1}{c|}{\NA} & .96$\uparrow$ & .99$\uparrow$ \\
  & Mitigation & .26$\uparrow$ & .31$\uparrow$ & \multicolumn{1}{c}{\NA} & \multicolumn{1}{c|}{\NA} & .50$\uparrow$ & .55$\uparrow$ \\
  \hline
 \multirow{2}{*}{Untargeted} & Vulnerability &  .34$\downarrow$ & .22$\downarrow$ & .47$\downarrow$ & .44$\downarrow$ & .51$\downarrow$ & .70$\downarrow$\\
 & Mitigation & .27$\downarrow$ & .21$\downarrow$ & .40$\downarrow$ & .54$\downarrow$ & .44$\downarrow$ & .62$\downarrow$ \\
 \hline
 \multicolumn{8}{c}{Embedding Dimensionality = 500, DNN Depth = 4} \\
  \hline
  \multirow{2}{*}{Targeted} & Vulnerability & .50$\uparrow$ & .56$\uparrow$ & \multicolumn{1}{c}{\NA} & \multicolumn{1}{c|}{\NA} & .91 $\uparrow$ & .95$\uparrow$ \\
 & Mitigation &  .27$\uparrow$ & .33$\uparrow$ & \multicolumn{1}{c}{\NA} & \multicolumn{1}{c|}{\NA} & .52$\uparrow$ & .54$\uparrow$ \\
  \hline
  \multirow{2}{*}{Untargeted} & Vulnerability &  .30$\downarrow$ & .19$\downarrow$ & .44$\downarrow$ & .36$\downarrow$ & .49$\downarrow$ & .63$\downarrow$\\
 & Mitigation & .22$\downarrow$ & .15$\downarrow$ & .37$\downarrow$ & .42$\downarrow$ & .40$\downarrow$ & .61$\downarrow$ \\
 \hline
 \end{tabular}
\caption{Attack performance under alternative surrogate models.\label{tab:cyber:surrogate}}}
\end{table}

\vspace{2pt}
\textbf{Surrogate models --} Thus far, we assume the surrogate models on which the adversary crafts poisoning facts and/or adversarial queries share the same architectures with the target \krl system. We now examine the scenario wherein the surrogate and actual models differ. 

We consider two configurations of the surrogate models different from the actual models used by \krl ({\em c.f.} Table\mref{tab:model}) in terms of \mcounter{i} the dimensionality of embeddings and \mcounter{ii} the depth of \dnns (as relation transformation models), while all the other settings are the same as Table\mref{tab:cyber:attack:overall}. Table\mref{tab:cyber:surrogate} shows the attack performance under such alternative surrogate models.

\underline{{\em \system transfers across different embedding models.}} -- By comparing Table\mref{tab:cyber:surrogate} and Table\mref{tab:cyber:attack:overall}, it is observed that while the attack performance drops under the alternative settings, due to the discrepancy between the actual and surrogate models, the decrease is marginal, indicating the transferability of \system across different models. This may be explained by that many \kg embedding methods demonstrate fairly similar behavior\mcite{bilinear-embedding}. It is therefore possible to transfer attacks across different embedding models.

% Table\mref{tab:cyber:surrogate} shows overall results when the surrogate system is different with the actual one. Comparing with default model configurations in Table\mref{tab:model}, our surrogate models take two group of alternative structure on both embedding dimension and the DNN depth. One group of surrogate configuration has less dimension and shallow depth of reasoning model (dim=200, depth=1), while the other has larger dimension and deeper architecture (dim=200, depth=4). Note that, both settings result in worse attack performances ({\em cf.} Table\mref{tab:cyber:attack:overall} due to the deviate models structure, hence do not fully transfer the attack ability from the surrogate system to the actual one.

\vspace{2pt}
\textbf{KG-Query interaction --} We further evaluate the interaction between the attack vectors of knowledge poisoning and query perturbation in \aco. Specifically, we evaluate the attack performance as a function of $n_\mathrm{kp}$ (number of injected poisoning facts) and $n_\mathrm{qp}$ (number of perturbed logical paths), with results summarized in Figure\mref{fig:cyber:cop:ndcg}.

\underline{{\em There exists an ``mutual reinforcement'' effect.}} -- In both targeted and untargeted cases, with $n_\mathrm{qp}$ fixed, slightly increasing $n_\mathrm{kp}$ significantly improves the attack performance. For instance, in targeted cases, when $n_\mathrm{kp} = 0$, \ndcgf remains 0 regardless of the setting of $n_\mathrm{kp}$; if $n_\mathrm{kp} = 50$, even setting $n_\mathrm{qp} = 1$ leads to \ndcgf over 0.5. This indicates that knowledge poisoning greatly boosts the effectiveness of query perturbation and also validates the analysis in \msec{sec:co}.

\underline{{\em The effect is even more evident in untargeted attacks.}} -- It is also observed that \system achieves its maximum effectiveness in untargeted attacks with $n_\mathrm{kp}$ and $n_\mathrm{qp}$ lower than the case of targeted attacks. This is explained by that without the need to direct \krl to a specific answer $A$, untargeted attacks are relatively ``easier'' than targeted attacks, making the mutual reinforcement even more significant.

% Besides above evaluations on the default setting, we also adjust two-factor constraints, \mie, poisoning budget $\gM$ and evasion scale $\gS$, to see variations of \cop attack, and consider alternative $\ssub{\gL}{tar}$ with different lengths.

% \cop is constrained by both the poisoning budget $\gM$ and the evasion scale $\gS$ to ensure evasiveness. We are curious about how those constraints affect the overall attack efficacies and what is the trend. Figure\mref{fig:cyber:cop:ndcg} shows the performance ({\it NDCG@5}) variation of \cop when adjusting both $\gM$ and $\gS$. 

% In targeted inference, we observe that when $\gM=0$, the NDCG@5 stays 0 even increasing $\gS=4$. As long as $\gM>0$ (\meg, $\gM=50$), relaxing $\gS$ from 0 to 1 significantly increases the attack performance. While $\gS=2\rightarrow4$ makes \cop more effective, attacking the mitigation-query application has more increments than attacking the threat-code-query ({\em cf.} Figure\mref{fig:cyber:cop:ndcg}-(a) \& (b)), in which \cop on the threat-code-query application is easier to reach to a stabilization.

% In general degradation, soley relaxing $\gS$ achieves bigger increments than solely relaxing $\gM$. Compared with targeted-inference results, \cop is easier to reach to the optimal efficacy (\mie, yellow region) and stabilize its performance, hence demonstrating that the general degradation is an easier-to-achieve objective.

We also observe similar trends measured by \mrr with results shown in Figure\mref{fig:cyber:cop:mrr} (\msec{sec:additional}).

\vspace{2pt}
\textbf{Alternative $p^*$ pattern --} The pattern $p^*$ serves a trigger to invoke \krl to malfunction in \akp and \aco. Here, we consider alternative definitions of $p^*$ and evaluate its impact on the attack performance. Specifically, besides its default definition (with $\textsf{\small Android OS}$ as anchor) in \msec{sec:case1:attack}, we consider two other definitions as listed in Table\mref{tab:diff:tar:logic}: the one has $\textsf{\small Port Scanning}$ (which refers to an {\sl attack pattern}) as its anchor and its logical path is of length 2; the other has $\textsf{\small T1033}$ (which refers to an {\sl attack technique} of ``system owner/user discovery technique''\footnote{https://attack.mitre.org/techniques/T1033/}) as its anchor and is of length 3.

% \kgp relies on falsifying the embedding semantics of $\ssub{\gL}{tar}$ to affect reasoning results. Hence the selection of $\ssub{\gL}{tar}$ affects both $SA_{kgp}$ and \cop attacks that involved with \kgp. Here we consider alternative $\ssub{\gL}{tar}$ that not only differentiate in the category but vary in length. Table\mref{tab:diff:tar:logic} contains information of our choices: by default, our $\ssub{\gL}{tar}$ starts with the anchor entity $\mathsf{Android\,\,OS}$, which is a product entity in our CyberKG. Besides, we also consider the anchor entity in  $\ssub{\gL}{tar}$ as an attack-pattern term named $\mathsf{Port\,\,Scanning}$ and a threat-technique code $\mathsf{T1033}$ (refers to a $\mathsf{System\,\,Owner/User\,\,Discovery}$ technique\footnote{https://attack.mitre.org/techniques/T1033/}). All three choices have diverse categories in their anchor entities and different length of relational paths.

\begin{table}[!ht]
\renewcommand{\arraystretch}{1.2}
\centering
\setlength{\tabcolsep}{2.0pt}
{\footnotesize
\begin{tabular}{ c|ccc}
Anchor of $p^*$ & $\mathsf{Android\,\,OS}$ & $\mathsf{Port\,\,Scanning}$ & $\mathsf{T1033}$\\
\hline
Category & $\mathsf{Product}$ & $\mathsf{Attack\,\,pattern}$ & $\mathsf{Technique}$ \\
Length (Vulnerability) & 1 hop & 2 hop & 3 hop \\
Length (Mitigation) & 2 hop & 3 hop & 4 hop \\
\end{tabular}
\caption{Alternative definitions of $p^*$. \label{tab:diff:tar:logic}}}
\end{table}

\underline{{\em $p^*$ with shorter paths leads to more effective attacks.}} -- In Figure\mref{fig:cyber:diff:tar:logic}, we compares the attack performance of \akp and \aco under varying definitions of $p^*$ in both targeted and untargeted cases. 
%with different $\ssub{\gL}{tar}$. The $x$-axis from left to right are length $\ssub{\gL}{tar}$ in each specific application (\mie, querying threat code/mitigation). We do not include $SA_{lqe}$ since it is not affected by the choice of $\ssub{\gL}{tar}$. 
Observe that the attack effectiveness decreases as the length of $p^*$ grows. This can be explained as follows. As the poisoning facts are selected surrounding the anchors, their influence tends to fade as the distance between the anchors and the answers. Further, while \akp is inefficient under the setting of $p^*$ starting with $\textsf{\small T1033}$, \aco is much less sensitive to the setting of $p^*$ (\meg, \mrr $\geq 0.39$ and \ndcgf $\geq 0.42$), indicating the necessity of co-optimizing poisoning facts and adversarial queries.

% because the poisoning effect is harder to retain in query embedding if the perturbed anchor entity locates in a long path. Without auxiliary of \lqe, the $SA_{kgp}$ almost loses all its efficacy when $\ssub{\gL}{tar}$ starts with $\mathsf{T1033}$, while \cop is still effective ({\it MRR}$\geq$0.39, {\it NDCG@5}$\geq$0.42)  in both objectives, which further demonstrates the necessity of the co-optimized strategy.

\vspace{2pt}
{\bf Zero-day threats --} We further evaluate the performance of \system in the scenario of {\it zero-day} threats, wherein the threats exploit previously unknown vulnerabilities, which often result in consequential damages in practice\mcite{zero-day}.  
We simulate the zero-day case by randomly removing 30\% \cves (from 2020 to 2021) from the \kg as the zero-day vulnerabilities, then generating queries with ground-truth answers among the deleted \cves. The reasoning task is to search for potential {\sl mitigation} for given incidents. Intuitively, it is possible to find correct mitigation even though the vulnerability is unknown based on similar vulnerabilities (\mie, approximate query). Meanwhile, the attacks aims to mislead \krl to erroneous suggestions (untargeted) or a specific {\sl mitigation} $A = \textsf{{\small MITI-72591}}$\footnote{https://source.android.com/security/bulletin/2021-04-01} (targeted).

\vspace{2pt}
\begin{table}[!ht]
\renewcommand{\arraystretch}{1.2}
\centering
\setlength{\tabcolsep}{3pt}
{\footnotesize
\begin{tabular}{ c|lr|lr|lr|lr}

 \multirow{2}{*}{Objective} & \multicolumn{8}{c}{Attack} \\ 
 \cline{2-9} 
  & \multicolumn{2}{c|}{w/o} & \multicolumn{2}{c|}{\akp} & \multicolumn{2}{c|}{\aqp} & \multicolumn{2}{c}{\aco} \\
  \hline
  \hline
 Targeted &  .00 & .00 & .34$\uparrow$ & .40$\uparrow$ & \multicolumn{1}{c}{\NA} & \multicolumn{1}{c|}{\NA} & .47$\uparrow$ & .55$\uparrow$ \\
 Untargeted &  .39 & .61 & .20$\downarrow$ & .22$\downarrow$ & .30$\downarrow$ & .44$\downarrow$ & .39$\downarrow$ & .59$\downarrow$ \\
 \hline
 \end{tabular}
\caption{Attack performance against zero-day threat queries. \label{tab:cyber:attack:zeroday}}}
\end{table}

\underline{{\em \system is also effective against approximate queries.}} -- Table\mref{tab:cyber:attack:zeroday} presents the attack performance in zero-day case. Comparing with Table\mref{tab:cyber:attack:overall}, we have similar observations that ({\em i}) \akp is more effective than \aqp in targeted attacks; ({\em ii}) \aqp is more effective than \akp in untargeted attacks; and ({\em iii}) \aco is the most effective among the three.

Also, note that \system is marginally less effective against approximate queries, due to the missing links (\mie, {\sl Vulnerability}) on the paths from anchor entities to answers ({\sl Mitigation}). However, as similar vulnerabilities tend to share mitigation, \system is still able to craft effective poisoning facts/adversarial queries to mislead \krl.

% because zero-day threat entities impede the transfer of poisoning effects. A CyberKG with zero-day threats has no direct path from anchor entities to ground-truth mitigation codes. Hence the reasoning process can only find those mitigation codes based on other similar threat codes. This prevents attacks from correctly matching the expected threat codes then finding mitigation results in the embedding space.

\section{Case Study II: Drug Repurposing}
\label{sec:drug:case}

In the case of cyber-threat hunting, we evaluate \system over queries regarding \kg entities. Below, we further evaluate \system over queries regarding \kg relations in the concrete case of drug repurposing reasoning. 

\subsection{Experimental setup}

We begin by describing the evaluation setting.

\vspace{2pt}
{\bf Scenario --} Drug repurposing aims to find potential therapeutic values or unknown side effects of new drugs\mcite{pushpakom2019drug, oprea2011drug} by exploring the interactions between drugs, diseases, and human genes. To this end, we consider a \krl system built upon knowledge regarding drugs, diseases, and genes, which supports the following queries:

% With the auxiliary of reasoning techniques, researchers can more efficiently explore interactions between drugs and their targets, such as diseases or genes. Even though the reasoning results still need to be verified by biomedical tests, the reasoning-based approaches significantly save efforts from enumerating all repurposing candidates.

% Here we consider two applications to address the drug repurposing task: 

{\em Drug-disease interaction --} Given a drug-disease pair  with their relevant properties, it determines the drug's effect on the disease (\meg, highly/mildly effective).
% we aim to explore how the drug interacts with this disease.

{\em Drug-gene interactions --} Given a drug-gene pair  with their relevant properties, it determines the gene's effect on the drug's functionality (\meg, suppression/acceleration). 

Both types of queries can be formulated as reasoning about the relation $r_? \in \gR$ between a given pair of \kg entities $v, v'$: $q[r_?] = r_? \,.\, \exists v \xrightarrow{r_?} v'$.

% In summary, we use the ontological reasoning mechanism in cyber threat intelligence and aim to find the threat codes and mitigation entities with given evidence. We treat the drug repurposing task as a relational reasoning problem, where we focus on inferring correct interactions between drugs and their targets despite other biomedical use cases\mcite{zhu2020knowledge, zhang2021drug}.

%\vspace{2pt}

\vspace{2pt}
{\bf DRKG -- } We use the Drug Repurposing Knowledge Graph ({\small DRKG})\mcite{drkg2020} as the underlying \kg, which is synthesized from public medical databases. Table\mref{tab:drkg:stat} lists the stats of {\small DRKG}. In particular, there are 10 different drug-disease relations and 34 drug-gene relations. 

\vspace{2pt}
{\bf Queries --} To construct the queries, we randomly sample 80K drug-disease and 200K drug-gene pairs and also include their 2-hop neighboring entities as their relevant properties. We use 5\% of the queries as the testing set and the remaining as the training set for \krl.  

%Our queries are subgraphs centralized by a drug-target pair, where the target is either a disease or a gene entity. Each query also contains 2-hop neighbor entities as properties of drugs/targets. We extract all drug-disease/gene pairs and their 2-hop surrounding subgraphs, then use 5\% as test queries and the rest as train queries. 

\vspace{2pt}
{\bf KRL --} We instantiate {\small GQE}\mcite{logic-query-embedding} as the encoders and {\small R-GCN}\mcite{rgcn} as the operators of the \krl system. Intuitively, {\small R-GCN} aggregates multi-relational graphs with {\small GCN} layers\mcite{gcn}, thereby able to aggregate query embeddings to predict missing relations.

\begin{table}[!ht]
\renewcommand{\arraystretch}{1.2}
\centering
{\small
 \begin{tabular}{c|c}
%  \toprule
% \Xhline{1.5\arrayrulewidth}
 Query &  HIT@1 \\ 
\hline 
\hline
Drug-Disease & 0.73 \\ 
Drug-Gene & 0.77 \\
 \end{tabular}
 
\caption{\krl performance in drug repurposing. \label{tab:benign:drug}}}
\end{table}

\vspace{2pt}
{\bf Metrics --} As the answer of a relation query is given as the ranking of possible relations, in the evaluation, we mainly use \hit ($K = 1$ by default) as the metric: for each query, it checks whether the top-$K$ answers contain the ground-truth answer (\mie, a binary indicator). The results are then averaged across all the queries. Table\mref{tab:benign:drug} shows the performance of \krl in this case.

\begin{figure*}
    \centering
    \epsfig{file = 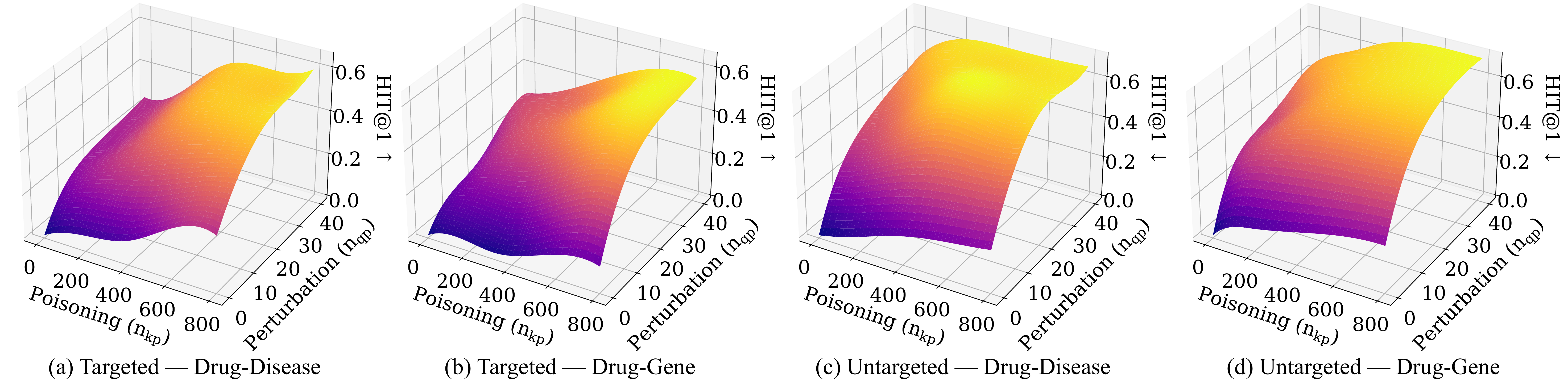, width = 170mm}
    \caption{HIT@1 variation of \aco when adjusting \kg poisoning budget ($\mathrm{n}_\mathrm{kp}$) and query perturbation budget ($\mathrm{n}_\mathrm{qp}$)}
    \label{fig:drug:cop}
\end{figure*}

\subsection{Attack implementation}

Next, we detail the implementation of \system in drug repurposing reasoning. 

\vspace{2pt}
{\bf Attack settings --} By default, we set target queries $\gQ^*$ as that contains trigger 
$p^* = \textsf{\small Nervous System}\xrightarrow[]{\text{includes}} v_\mathrm{drug}$,
where $\textsf{\small Nervous System}$ is an Anatomical Therapeutic Chemical ({\small ATC}) category as the anchor and $v_\mathrm{drug}$ is a variable. Intuitively, the adversary aims to influence the reasoning regarding all the drugs in this {\small ATC} category.

% The relational repurposing between a drug to its targets (diseases, genes) emphasizes drug entities, hence the targeted logic $\ssub{\gL}{tar}$ should relate to a set of drugs. Based on the drug repurposing \kg, we pick a specific $\mathsf{Atc}$ entity as anchor node and its relation to drug entities as the logical path in $\ssub{\gL}{tar}$. $\mathsf{Atc}$ or Anatomical Therapeutic Chemical \footnote{https://go.drugbank.com/atc} is a classification system among drugs based on their ingredients\mcite{}\ting{cite ATC papers from wikipedia}. We aim to affect a set of drugs belong to a targeted $\mathsf{Atc}$ class, then influence the reasoned relation.

We consider both targeted and untargeted attacks. In targeted cases, the goal is to direct the answering each query $q \in \gQ^*$ to a specific relation $A = \textsf{\small biomarker-of}$; in untargeted cases, the goal is to deviate the answering of $q$ from its ground-truth answer $\llbracket q \rrbracket$.

% Attack objectives also focus on the reasoned relation: the targeted inference picks a non-ground-truth relation between a drug and its targets and puts efforts into making all reasoning results as the targeted relation as long as the drug belongs to the targeted $\mathsf{Atc}$ class; the general degradation only concerns whether the reasoning results are the erroneous w.r.t. drugs belongs to our targeted $\mathsf{Atc}$.

\vspace{2pt}
{\bf Attack variants --} We compare the performance of different variants of \system. 

%\vspace{1pt}
\akp -- It influences \krl by injecting poisoning facts to the \kg. Given that genes and diseases are public, verifiable knowledge, it is more evasive to inject poisoning facts surrounding drugs. Thus, \akp attaches poisoning facts to $\textsf{\small Nervous System}$ ($p^*$'s anchor). 
% The \underline{s}tand\underline{a}lone \kgp attack. The attacker is only accessible to \kg as a malicious contributor. A poisoned \kg affects downstream embedding and reasoning without further manipulation of the \kgp attacker.

%\vspace{1pt}
\aqp -- It directly attaches additional logical paths (from the \kg) to the drug entity of the query. As all the related properties of a drug are within 2 hops, \aqp also searches for logical paths within 2 hops of the drug.

% which simulates a practical issue that malicious properties are provided on a victim drug. Since the drug properties are 2-hop away at most, \lqe also searches for paths no longer than 2-hop as perturbations.
% Relying on query perturbation, the adversary generates adversarial queries at inference time, but has no control over the \kg construction. To make the perturbation evasive, we require that \mcounter{i} the number of logical paths in the perturbation is limited by  $n_\mathrm{qp} = 2$ and \mcounter{ii} the dependency graph of the perturbed query is valid with respect to the \kg. 

%The \underline{s}tand\underline{a}lone \lqe attack. The attacker locates at downstream use cases, thus only affects query contents, including anchor entities and logical structures.

%\vspace{1pt}
\aco  -- Leveraging both knowledge poisoning and query perturbation, \aco optimizes the poisoning facts and adversarial queries jointly. 

Given that {\small DRKG} (average density 120.8) is much denser than {\small CyberKG} (average density 8.7), we limit the number of injected poisoning facts by $n_\mathrm{kp} = 400$ and the number of perturbed logical paths $n_\mathrm{qp} = 20$ by default. The setting of other parameters is deferred to Table\mref{tab:model}.

\begin{figure*}
    \centering
    \epsfig{file = 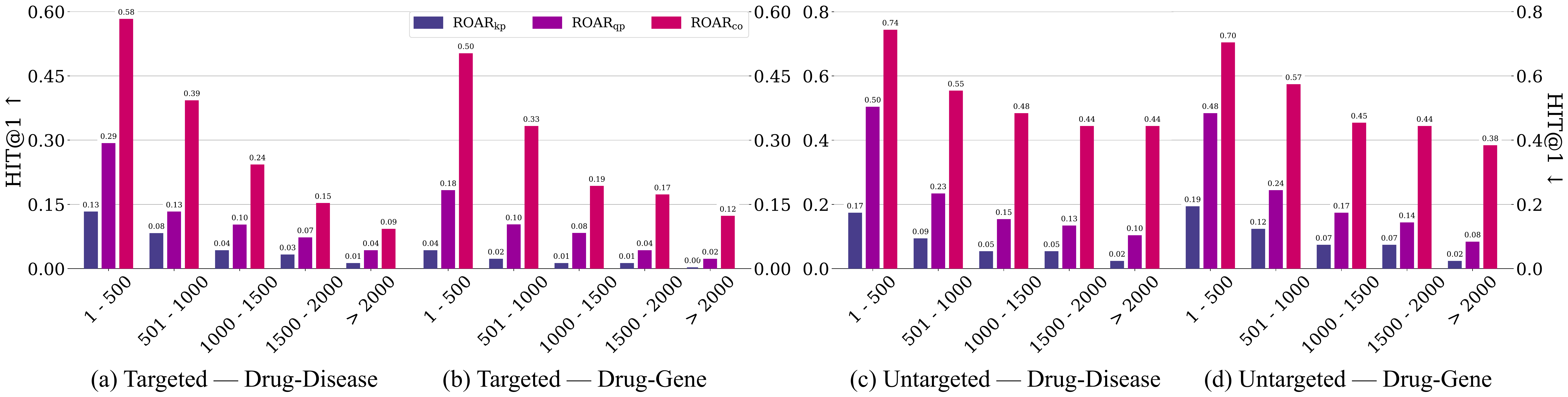, width = 172mm}
    \caption{Attack performance with respect to the number of properties associated with drug entities.}
    \label{fig:drug:diff:dense}
\end{figure*}

\subsection{Evaluation results}

The evaluation below focuses on three aspects: ({\em i}) the overall attack effectiveness and evasiveness ({\em ii}) the interaction between the two attack vectors, and ({\em iii}) the impact of different factors on the attack performance.

\begin{table}[!ht]
\renewcommand{\arraystretch}{1.2}
\centering
\setlength{\tabcolsep}{1pt}
{\footnotesize
\begin{tabular}{ c|c|cccc|cc}
 \multirow{2}{*}{Objective} & \multirow{2}{*}{Query} &  \multicolumn{4}{c|}{Target $\gQ^*$} &  \multicolumn{2}{c}{Non-Target $\gQ \setminus \gQ^*$}\\ 
 \cline{3-8} 
  & & w/o & \akp & \aqp & \aco & \akp & \aco \\
  \hline
  \hline
 \multirow{2}{*}{Targeted} & Drug-Disease & .00 & .10$\uparrow$ & .22$\uparrow$ & .46$\uparrow$ &  .03$\downarrow$ & .03$\downarrow$ \\
  &  Drug-Gene & .04 & .03$\uparrow$ & .14$\uparrow$ & .40$\uparrow$ & .01$\downarrow$ & .00$\downarrow$ \\
  \hline
 \multirow{2}{*}{Untargeted} & Drug-Disease & .68 & .13$\downarrow$ & .38$\downarrow$ & .65$\downarrow$ & .02$\downarrow$ & .03$\downarrow$ \\
  & Drug-Gene & .74 & .15$\downarrow$ & .37$\downarrow$ & .62$\downarrow$ & .03$\downarrow$ & .01$\downarrow$ \\
 \hline
 \end{tabular}
\caption{Overall attack performance of \system in drug repurposing. \label{tab:attack:drug}}}
\end{table}

\vspace{2pt}
{\bf Attack effectiveness --} Table\mref{tab:attack:drug} presents the overall performance of \system measured by \hito. Similar to the case of cyber-threat hunting, observe that \aco outperforms the other variants in both targeted and untargeted cases, due to the interactions between knowledge poisoning and query perturbation. Besides, we have the following observation:

\underline{\textit{\aqp is more effective than \akp.}} -- Across all the settings, \aqp consistently outperforms \akp. This may be explained as follows: \akp influences the drug-disease (or drug-gene)  relation by attaching poisoning facts to the {\small ATC} entity, which in turn influences the drug entity; meanwhile, \akp directly attaches perturbation to the drug entity, leading to more effective attacks.

\vspace{2pt}
{\bf Attack evasiveness --} Table\mref{tab:attack:drug} also shows the impact of \system on non-target queries $\gQ \setminus \gQ^*$. Observe that the \hito score drops no more than 0.03 across all the settings, indicating the fairly marginal impact of \system.

% The {\it HIT@1} drops no more than 0.03, hence ensures that \system incurs ignorable side effects to normal reasoning ability and demonstrates the evasiveness of our attacks. This is similar to effects on the threat engine ({\em cf.} Table \mref{tab:benign:drop:cyber}).

\vspace{2pt}
\textbf{KG-Query interaction --} We further evaluate the interaction between knowledge poisoning and query perturbation in the case of drug repurposing. We evaluate the attack performance as a function of $n_\mathrm{kp}$ and $n_\mathrm{qp}$, with results summarized in Figure\mref{fig:drug:cop}.

Similar to \msec{sec:cyber:case}, we have the observations below. \mcounter{i} There exists an ``mutual reinforcement'' effect between the two attack vectors. The injected poisoning facts greatly boots the effectiveness of adversarial queries in directing the answering of relation queries. \mcounter{ii} This reinforcement effect seems more evident in untargeted attacks, given that an untargeted attack succeeds if the drug-disease (or drug-gene) relations are mispredicted while a targeted attack requires the relation to be predicted to target answer $A$.

\begin{table}[!ht]
\renewcommand{\arraystretch}{1.2}
\centering
\setlength{\tabcolsep}{0.5pt}
{\footnotesize
\begin{tabular}{ c|c|ccc|ccc}
 \multirow{2}{*}{Objective} & \multirow{2}{*}{Query} &  \multicolumn{3}{c|}{\#Dim = 100, \#Layer = 1} &  \multicolumn{3}{c}{\#Dim = 400, \#Layer = 4}\\ 
 \cline{3-8} 
  & & \akp & \aqp & \aco & \akp & \aqp & \aco \\
  \hline
  \hline
 \multirow{2}{*}{Targeted} & Drug-Disease & .10$\uparrow$ & .18$\uparrow$ & .41$\uparrow$ & .05$\uparrow$ & .12$\uparrow$ & .33$\uparrow$ \\
  & Drug-Gene & .01$\uparrow$ & .09$\uparrow$ & .34$\uparrow$ & .00$\uparrow$ & .12$\downarrow$ & .30$\downarrow$ \\
  \hline
 \multirow{2}{*}{Untargeted} & Drug-Disease & .11$\downarrow$ & .35$\downarrow$ & .62$\downarrow$ & .10$\downarrow$ & .24$\downarrow$ & .51$\downarrow$ \\
  & Drug-Gene &  .15$\downarrow$ & .33$\downarrow$ & .57$\downarrow$ & .06$\downarrow$ & .28$\downarrow$ & .49$\downarrow$\\
 \hline
 \end{tabular}
\caption{Attack performance under alternative surrogate models in drug repurposing. \label{tab:drug:surrogate}}}
\end{table}

\vspace{2pt}
{\bf Surrogate models --} We now evaluate \system in the case wherein the surrogate and actual models differ. We consider two configurations of the surrogate models different from the actual models used by \krl ({\em c.f.} Table\mref{tab:model}) in terms of \mcounter{i} the embedding dimensionality and \mcounter{ii} the depth of {\small R-GCN}, while all the other settings are the same as Table\mref{tab:attack:drug}. 

It is shown in Table\mref{tab:drug:surrogate} that while the attack performance slightly decreases as the surrogate and actual models differ, the drop is marginal, especially when the surrogate model is simpler than the actual model (embedding dimensionality = 100, depth of {\small R-GCN} = 1), indicating the transferability of \system from simpler models to more complicated ones.

% \underline{{\em \system transfers across different embedding models.}} -- By comparing Table\mref{tab:cyber:surrogate} and Table\mref{tab:cyber:attack:overall}, it is observed that while the attack performance drops under the alternative settings, due to the discrepancy between the actual and surrogate models, the decrease is marginal, indicating the transferability of \system across different models. This may be explained by that many \kg embedding methods demonstrate fairly similar behavior\mcite{bilinear-embedding}. It is therefore possible to transfer attacks across different embedding models.

\vspace{2pt}
{\bf Number of properties --} Recall that a relation query is a subgraph centering around a drug-disease (or drug-gene) pair, while the surrounding entities represent their descriptive properties. For instance, the neighboring entities of a drug entity describe its biomedical properties (\meg, {\sl ATC} and {\sl side effect}). Figure\mref{fig:drug:diff:dense} breaks down the attack performance according to the number of properties associated with drug entities.  

% A central node with denser properties brings more difficulty to attacks. We are curious how does \system vary its performances with different drug-property densities. 

% Figure \mref{fig:drug:diff:dense} presents attack performances with different density levels of drug entities. 

\underline{\em Relations with fewer properties are more vulnerable.} -- It is observed that in both targeted and untargeted cases, the attack performance degrades with the number of properties. This can be explained as follows. For given query $q$, each property represents one logical constraint on its ground-truth answer $\llbracket q \rrbracket$; more properties make \krl more robust against poisoning facts or adversarial perturbation. Yet, even with more than 2,000 properties, \aco still attains high \hito scores, especially in the untargeted cases, indicating the necessity of optimizing poisoning facts and adversarial queries jointly.

% We split density ranges from 0-500 to $>$2000 and observe that the increasing density impedes all attack efficacies with different trends. However, even though the attack efficacies significantly diminish when density ranges from 0 to 1500, we observe a stabilization as density $>$1500. The $SA_{lqe}$ and \cop is even successful when density $>$2000. Recap that the evasion budget $\gS=20$ by default, which only occupies less than 1\% of surrounding nodes, demonstrates that the evasion attack resists its efficacy even when a drug has thousand-level properties.
\section{Discussion}
\label{sec:discussion}

We now explore potential countermeasures against \system.

\subsection{Potential Countermeasures}

Due to the unique characteristics of \krl, the existing defenses against malicious attacks in classification tasks\mcite{pgd,neural-cleanse,spectral-signature}) are inapplicable. Thus, we investigate two potential countermeasures tailored to knowledge poisoning and query perturbation, respectively, and further explore their synergy.

% \system represents a new class of attacks on reasoning systems; hence existing mitigation against adversarial vulnerabilities\mcite{pgd, adv-train-free, zhang2019defense} or backdoors\mcite{Wang:2019:sp, Gao:2019:arxiv, abs} are inapplicable. In this part, we adopt separate countermeasures to remedy adversarial effects on \kg and use-case queries, respectively, which are: ({\em i}) a subsampling-based embedding that removes potential \kg poisoning; ({\em ii}) robust training that prevents downstream evasions.  

\vspace{2pt}
{\bf Filtering poisoning facts --} As poisoning facts are forcefully injected into the \kg, they may misalign with their neighboring entities/relations. It is therefore possible to detect and purge them as noisy facts before using the \kg in \krl.

To this end, for each fact $v \xrightarrow{r} v'$ in the \kg, we apply encoder $\phi$ and relation $r$-specific operator $\psi_r$ to assess its ``fitness'' to the \kg. Specifically, we compute the anomaly score of $v \xrightarrow{r} v'$ as: $\| \psi_r(\phi_v) - \phi_{v'} \|$. A higher anomaly score indicates less fitness of $v \xrightarrow{r} v'$. We may remove facts with the highest scores, thereby mitigating the influence of poisoning facts.

In practice, we may first train \krl (including both $\phi$ and $\psi$) with the complete \kg, prune $m\%$ of the facts with the highest anomaly scores, and then re-train \krl. 

% Poisoning facts affect local \kg structures, falsifies the existing dependencies, and worsens corresponding embedding semantics. Therefore, we can describe the poisoned facts as {\it noisy facts}, which lead to a poor embedding score measured by $\psi$. Even though the defender (\mie, the reasoning system developer) is unaware of whether poisoned facts exist, she can enhance embedding robustness by removing those {\it noisy facts}.

% Recap that for each \kg fact $\left<u, r, v\right>$, the \kg embedding relies on a relation-specific function $\ssub{f}{r}$ to calculate a score $\ssub{f}{r}\left(u, v\right)$ with entity embeddings. We use a distance-based score in both cases; hence a larger value represents $u, v$ have a more disparate relation $r$ in embedding space, leading to worse embedding semantics. 

% Intuitively, poisoned facts are not naturally existed in \kg, hence perform as noisy. We can enhance the embedding robustness by subsampling \kg based on computed $\ssub{f}{r}$ scores. We remove facts with worse (larger) scores, thus mitigate effects from noisy facts.   

\begin{figure}[!ht]
    \centering
    \epsfig{file = 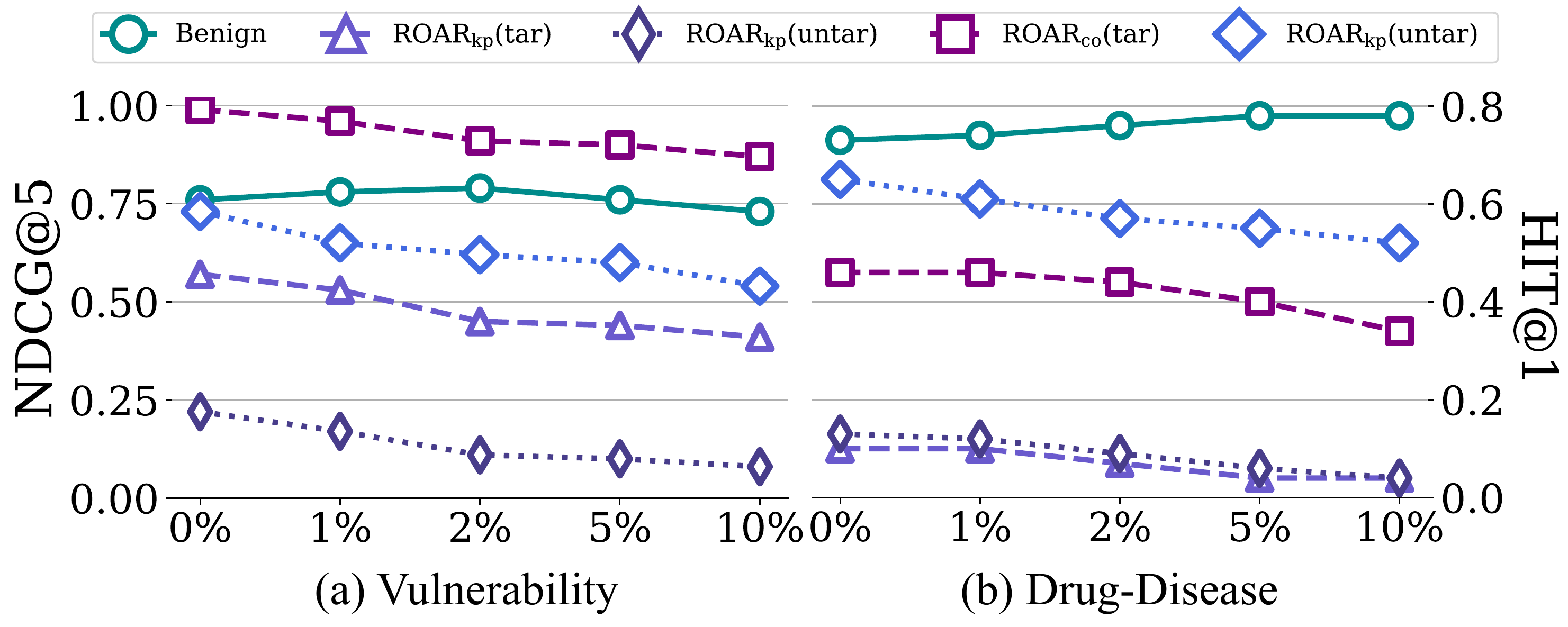, width = 84mm}
    \caption{\krl performance on non-target queries and attack performance on target queries. (a) Vulnerability query in cyber-threat hunting; (b) Drug-disease query in drug repurposing.}
    \label{fig:robust:kge}
\end{figure}

\begin{figure*}[!ht]
    \centering
    \epsfig{file = 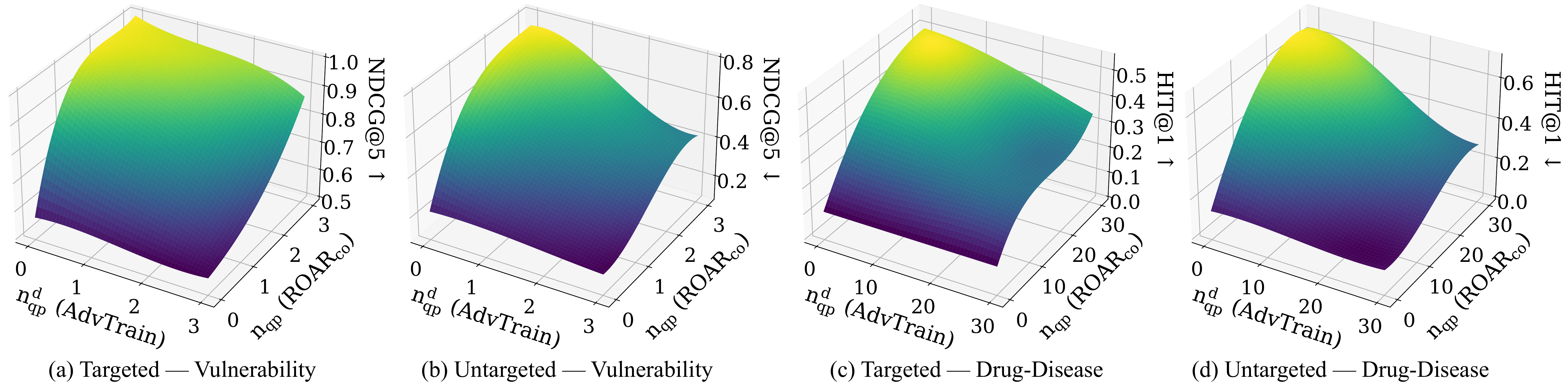, width = 170mm}
    \caption{Performance of \aco against robust training with respect to varying settings of $n_\mathrm{qp}$ and  $n_\mathrm{qp}^\mathrm{d}$.}
    \label{fig:adv:train}
\end{figure*}

{\em Results and analysis} -- Figure\mref{fig:robust:kge} shows the \krl performance on non-target queries and the attack performance on target queries in (a) cyber-threat hunting and (b) drug repurposing. It is observed that as $m$ grows from 0 to 10, in (a) the \krl performance slightly decreases while in (b) the \krl performance slightly improves. We attribute this phenomenon to the density difference of these two {\small KGs} (8.7 versus 120.8). For a sparser \kg like {\small CyberKG}, over-pruning has a negative impact on the \krl performance. In both cases, the attack performance slightly drops with $m$.

Thus, \mcounter{i} filtering of poisoning facts may not be suitable for sparse {\small KGs}; and \mcounter{ii} even with a large pruning rate (\meg, 10\%), the improvement of attack resilience seems marginal.

% and efficacies of $SA_{kgp}$ and $\cop$ (who use poisoned facts), under both attack objectives. Observe that as we increase the pruning ratio, the benign reasoning performance increases while all attacks are degraded. 

% However, there is a reverse trend on CyberKG, upon which the benign performance slightly decreases. We explain by the lower density of CyberKG: pruning too many facts (\mie, edges) from CyberKG damage the connectivity among entities (\mie, nodes), thus causes the worse embedding. On the contrary, the denser DRKG improves its embedding robustness while pruning more facts because it retains its connectivity during pruning.

% Even though this subsampling approach appears helpful, it cannot significantly degrade the attack efficacies to 0. Pruning too many facts (\meg, 10\%) also damages entity dependencies in \kg. Hence, there is a tradeoff between preventing attacks (enhancing robustness) and maintaining higher \kg completeness.

\vspace{2pt}
{\bf Training with adversarial queries --} We further consider an adversarial training\mcite{pgd} strategy to defend against query perturbation. Intuitively, during \krl training, we generate an adversarial version $q^*$ for each query $q$ using \aqp and add $(q^*, \llbracket q \rrbracket)$ to the training set, where $\llbracket q \rrbracket$ is $q$'s ground-truth answer. Note that here we use the untargeted \aqp (\meq{eq:untargetd}) to generate $q^*$.

% which leverages adversarial sample in training time to prevent downstream evasion. We select 10\% training queries and customize an evasion $\ssub{q}{lqe}$ for each query $q$ -- with the same technique as \lqe -- then replace $q$ by $\ssup{q}{P}=q\wedge\ssub{q}{lqe}$. Note that we do not change original answers; because we use evasion perturbations to simulate {\it noises} unexpectedly existing in queries.

% The training objective is different with \lqe ({\em cf.} Eq.\mref{eq:lqe:loss}). Here we aim to disparate embedding semantics by injecting $\ssub{q}{lqe}$:
% \begin{equation}
% \begin{split}
% \begin{gathered}
%     \ssub{\vect{q}}{\vect{lqe}} = \arg\max_{\ssub{\vect{q}}{\vect{lqe}}}
%     \mathsf{Dist}\left(\ssup{\vect{q}}{\vect{P}}, \vect{q}\right) \\
%     \text{s.t.}\quad\ssup{\vect{q}}{\vect{P}}=\ssub{\gamma}{\wedge}\left(\vect{q}, \ssub{\vect{q}}{\vect{lqe}}\right)
% \end{gathered}
% \end{split}
% \label{eq:adv:train}
% \end{equation}

% \noindent
% then use beam search to reconstruct $\ssub{q}{lqe}$ with $\ssub{\vect{q}}{\vect{lqe}}$.

{\em Results and analysis} -- Similar to \aqp, robust training has a threshold $n_\mathrm{qp}^\mathrm{d}$ that limits the number of perturbed paths. We measure the performance of \aco under varying settings of $n_\mathrm{qp}$ and $n_\mathrm{qp}^\mathrm{d}$, with results shown in Figure\mref{fig:adv:train}. 

Observe that across all the cases, robust training greatly reduces the attack performance, especially against untargeted attacks when $n_\mathrm{qp}^\mathrm{d} \geq n_\mathrm{qp}$. However, robust training also significantly impacts the \krl performance, resulting in over 0.19 \ndcgf drop in cyber-threat hunting and over 0.11 \hito drop in drug repurposing. Further, it is inherently ineffective against \akp, which does not rely on adversarial queries.

% robust training cannot prevent standalone \kgp ($\gS=0$) since \kgp does not rely on an evasion sample. Moreover, our evaluations show that robust training damages the benign reasoning abilities -- {\it NDCG@5} in the threat engine decreases 0.19; {\it HIT@1} in drug repurposing system decreases 0.11.

\vspace{2pt}
{\bf Synergy --} Finally, we integrate the two defenses above and explore their synergy. We set pruning rate $m = 1\%$ and $n_\mathrm{qp}^\mathrm{d} = 2/20$ for cyber-threat hunting/drug repurposing. The attacks follow the default setting in Table\mref{tab:model}.

% also use both subsampling-based embedding and robust training to evaluate how they mitigate \system attacks. We set the pruning ratio as 1\% for both cases and $\ssub{\gS}{def}=2/20$ in cyber intelligence/drug repurposing. Attacks are in default settings. Table\mref{tab:cm:both} shows the degradation value of attack efficacies. 

\begin{table}[!ht]
\renewcommand{\arraystretch}{1.2}
\centering
\setlength{\tabcolsep}{3pt}
{\footnotesize
\begin{tabular}{ c|c|ccc|c}
 \multirow{2}{*}{Query} & \multirow{2}{*}{Objective} & \multicolumn{3}{c|}{Target $\gQ^*$} & Non-Target\\ 
 \cline{3-5} 
  & & \akp & \aqp & \aco & $\gQ \setminus \gQ^*$ \\
  \hline
  \hline
  \multirow{2}{*}{Vulnerability} & Targeted & 0.05$\downarrow$ & \NA & 0.26$\downarrow$ & \multirow{2}{*}{0.02$\downarrow$} \\
   & Untargeted & 0.08$\downarrow$ & 0.20$\downarrow$ & 0.36$\downarrow$ & \\
  \hline
  \multirow{2}{*}{Drug-Disease} & Targeted & 0.02$\downarrow$ & 0.12$\downarrow$ & 0.15$\downarrow$ & \multirow{2}{*}{0.04$\uparrow$}\\
   & Untargeted & 0.06$\downarrow$ & 0.18$\downarrow$ & 0.30$\downarrow$ & \\
 \hline
 \end{tabular}
\caption{Attack performance and \krl performance against the integrated defense. \label{tab:cm:both}}}
\end{table}

{\em Results and analysis} -- The results are shown in Table\mref{tab:cm:both}. Comparing with the original attack performance  ({\em cf.} Table\mref{tab:cyber:attack:overall} and\mref{tab:attack:drug}), the integrated defense reduces the attack effectiveness by a large margin, even though it does not completely block \system; moreover, it alleviates the negative impact on the \krl performance and even improves the \hito score in drug repurposing! Thus, the integration of multiple defenses seems a promising direction worth further investigation. 

% collaborative defense cannot fully prevent \system attacks; the mitigation performs better on general degradation than on targeted inference. In addition, using both defenses relieves the damage to benign reasoning abilities compared to solely using robust training.

\subsection{Limitations}

We now discuss the limitations of this work.

\vspace{2pt}
{\bf Alternative reasoning tasks --} We mainly focus on reasoning tasks with one target answer (entity/relation), while there exist other reasoning tasks (\meg, path reasoning\mcite{path-reasoning} aims to find a path with given starting and end entities). Intuitively, \system is ineffective in such tasks as it requires knowledge about the logical path and then perturbs intermediate entities on the path. We consider exploring the vulnerability of alternative reasoning tasks as our ongoing research.

%So far, our studied reasoning framework only has one unit of variables, \mie, either entities or relations. In a more complex reasoning case, both entity and relation are variables and need to be found out as part of answers. One example is path reasoning \mcite{}, which aims to search a path with given starting and end entities. Intuitively, \system cannot realize an effective attack since it needs to have knowledge of potential paths and perturb all intermediate entities in \kg; also, it is hard to control the long-path contents by only perturbing query entities (the starting/end entity pair). Therefore, a stronger attack strategy is required for multi-unit reasoning like path reasoning.

\vspace{2pt}
{\bf Input-space attacks --} While \system directly operates on \kgs (or queries) in the logical space, there are scenarios in which \kgs (or queries) are extracted from real-world inputs. For instance, cyber-threat queries may be generated by software testing and inspection. In such scenarios, it requires the perturbation to \kgs (or queries) to be mapped to valid real-world inputs (\meg, software). While input-space attacks are an ongoing area of research\mcite{input-space-attack}, we consider realizing \system in the input space represents unique challenges.

% Even though we assume the query perturbation directly operates on a query, in reality, queries exist in some raw format. For instance, people collect cyber-domain queries by software testing and inspect instructions to collect the vulnerability properties and product information; in healthcare, people create queries from biomedical reports or hospital records. To ensure the created query contains perturbations, an adversary needs to access the raw information sources and transfers perturbations into the same format.  As for the text-based data, it is easier to craft malicious reports and inject; as for the software, the adversary needs to inject backdoor functionalities to induce the software to a specific behavior then be collected by victim users.

\vspace{2pt}
{\bf Integration of defenses --} While it is shown that integrating filtering poisoning knowledge and training with adversarial queries greatly improves the attack resilience, it is unclear how to optimally integrate such defenses to balance the factors of attack robustness (\meg, against adaptive attacks), impact on \krl performance, and training cost. We consider answering these questions critical for improving the robustness of \krl in a practical setting.

\section{Related Work}

Next, we survey the literature relevant to this work.

\vspace{2pt}
{\bf Knowledge representation learning --} Knowledge graphs (\kgs) represent valuable information sources in various domains\mcite{mittal2019cyber,zhu2020knowledge}. Recent years have witnessed significant progress in using machine learning to reason over \kgs. The existing work can be roughly classified into two categories.

% Knowledge graphs (KGs) collected from ample importance sources supply a wide variety of applications, including cyber security\mcite{mittal2019cyber, mittal2016cybertwitter, pingle2019relext, kiesling2019sepses}, medication\mcite{chen2019robustly, sheng2020dsqa, malik2020automated, zhu2020knowledge}, biochemstry\mcite{farazi2020knowledge}, finance\mcite{ruan2016building, fu2018stochastic}, and sociology\mcite{tchechmedjiev2019claimskg, nguyen2019social}. 

One line of work aims to develop effective \kg embeddings\mcite{bordes2013translating, wang2014knowledge, lin2015learning, nickel2016holographic, yang2014embedding} such that the semantics of \kg entities/relations are effectively captured by their latent representations for applications such as link prediction\mcite{chains-of-reasoning,path-query}. Another line of work focuses on directly making predictions about complex logical queries\mcite{logic-query-embedding, query2box, beta-embedding, ren2021lego} such as first-order conjunctive queries.
The \krl models considered in this paper belong to this category.

\vspace{2pt}
{\bf Machine learning security --} With their increasing use in security-sensitive domains, machine learning (\ml) models are becoming the targets for malicious attacks\mcite{Biggio:2018:pr}. A variety of attack vectors have been exploited: adversarial evasion crafts adversarial inputs to force the target model to malfunction\mcite{goodfellow:fsgm,carlini-attack}; model poisoning modifies the target model's behavior (\meg, performance drop) via polluting its training data\mcite{model-reuse}; backdoor injection creates a trojan model such that any trigger-embedded input is likely to be misclassified\mcite{trojannn}; and functionality stealing constructs a replicate model functionally similar to a victim model\mcite{knockoff-net}. 

In response, another line of work strives to improve the resilience of ML models against such attacks. For instance, against adversarial evasion, existing defenses explore new training strategies (\meg, adversarial training)\mcite{pgd} and detection mechanisms\mcite{Gehr:2018:sp}. Yet, such defenses often fail when facing even stronger attacks\mcite{Athalye:2018:icml,deepsec}, resulting in a constant arms race between the attackers and defenders.

\vspace{1pt}
Despite the intensive research on \krl and \ml security in parallel, the security of \krl is largely unexplored. This work represents an initial step to bridge this gap.

%Concurrent to this work, it is shown in\mcite{adversarial-robustness-arch} that \nas models tend to be more vulnerable to adversarial evasion, while our work differs in considering a variety of attacks beyond adversarial evasion, providing possible explanations for such vulnerability, and investigating potential mitigation.

\section{Conclusion}

This work represents an in-depth study on the security of knowledge representation learning (\krl). We present \system, a new class of attacks that instantiate a variety of threats to \krl. We demonstrate the practicality of \system in two representative security-sensitive applications, raising concerns about the current practice of training and using \krl. Moreover, we discuss potential mitigation, which might shed light on applying \krl in a more secure manner. 

%real-world threats lying in reasoning applications. Moreover, we explore potential mitigation against \system, addressing a solid stride toward the robust reasoning over knowledge graphs.

\clearpage

\bibliographystyle{plain}
\bibliography{bibs/reasoning.bib}

\appendix

\section{Notations}

Table\mref{tab:notations} summarizes the important notations.

\begin{table}[!ht]{\footnotesize
\centering
\renewcommand{\arraystretch}{1.2}
\begin{tabular}{c | l }

Notation & Definition \\
\hline
\hline
\multicolumn{2}{l}{{Data -- knowledge graph related}} \\
\hline
$\gG$ & a knowledge graph  \\

$\langle v, r, v' \rangle$ & a $KG$ fact from entity $v$ to $v'$ with relation $r$ \\
$\gN, \gE, \gR$ & entity, edge, and relation set of $\gG$\\
$\gF$ & all facts in $KG$ \\
$\gF^*$ & a fact set contributed by the attacker \\

\hline
\multicolumn{2}{l}{{Data -- query related}} \\
\hline
$q$, $Q$ & a single query; a query set \\
$\llbracket q \rrbracket$ & $q$'s ground-truth answer(s) \\
$A$ & the targeted answer \\
$p^*$ & the targeted logical path for attack \\
$\gK_q$ & anchor entities of query $q$ \\
$\gK^*$ & anchor entities of $p^*$ \\
$\gQ^*$ & a query set that each included query has $p^*$\\
$\gQ \setminus \gQ^*$ & a query set that each included query doesn't have $p^*$\\
$q^+$ & the generated perturbations on $q$ \\
$q^*$ & conjunction of $q$ and $q^+$, \mie, the perturbed query \\

\hline
\multicolumn{2}{l}{{Model or embedding related}} \\
\hline
$\gamma$ & the reasoning system \\
$\phi$ & the encoder for generating entity embeddings \\
$\psi$, $\psi_r$ & the (relation $r$-specific) operator\\
$\phi_{\gG}$ & embeddings of all $KG$ entities \\
$\phi_v$ & entity $v$'s embedding  \\
$\phi_{q}$ & $q$'s embedding \\
\hline
\multicolumn{2}{l}{{Other parameters}} \\
\hline
$n_\mathrm{kp}$ & \kg poisoning budget \\
$n_\mathrm{qp}$ & query perturbation budget \\
$n_\mathrm{qp}^\mathrm{d}$ & query perturbation budget used by the defender\\
$n_\mathrm{iter}$ & Number of iterations in co-optimization \\
\hline
\end{tabular}
\caption{Notations, definitions, and categories. \label{tab:notations}}}
\end{table}

\section{Attack Taxonomy}
\label{sec:app:taxonomy}

Attack$_1$, Attack$_2$, Attack$_3$ -- They use knowledge poisoning as the attack vector but with different knowledge of the encoder $\phi$ and the operator $\psi$. The adversary without knowledge of encoder $\phi$ builds a surrogate one with self-determined embedding dimensionality (Attack$_1$, Attack$_3$). The adversary without knowledge of operator $\psi$ crafts an independent model that may function differently (Attack$_1$, Attack$_2$).

Attack$_5$, Attack$_6$, Attack$_8$ -- They take query perturbation as the attack vector. If the adversary has no knowledge of the reasoning system (Attack$_5$) or has partial knowledge (Attack$_6$), it must leverage a surrogate system or partial component for inference-time perturbation. With white-box access (Attack$_8$), the adversary can directly query the actual system and craft perturbations.

Attack$_9$, Attack$_{11}$, Attack$_{12}$ -- They operate on both \kg and queries as attack vectors. The major difference is leveraging an entire surrogate system (Attack$_8$), or a  surrogate encoder $\phi$ with the actual reasoning operator $\psi$ (Attack$_{11}$), or white-box access to both $\phi$ and $\psi$  (Attack$_{12}$) and manipulate poisoning facts with query perturbations.

\section{KG Statistics}
\label{sec:kg:info}

Here, we present the statistics of \kgs used in \msec{sec:cyber:case} and \msec{sec:drug:case}.

\begin{figure}
    \centering
    \epsfig{file = 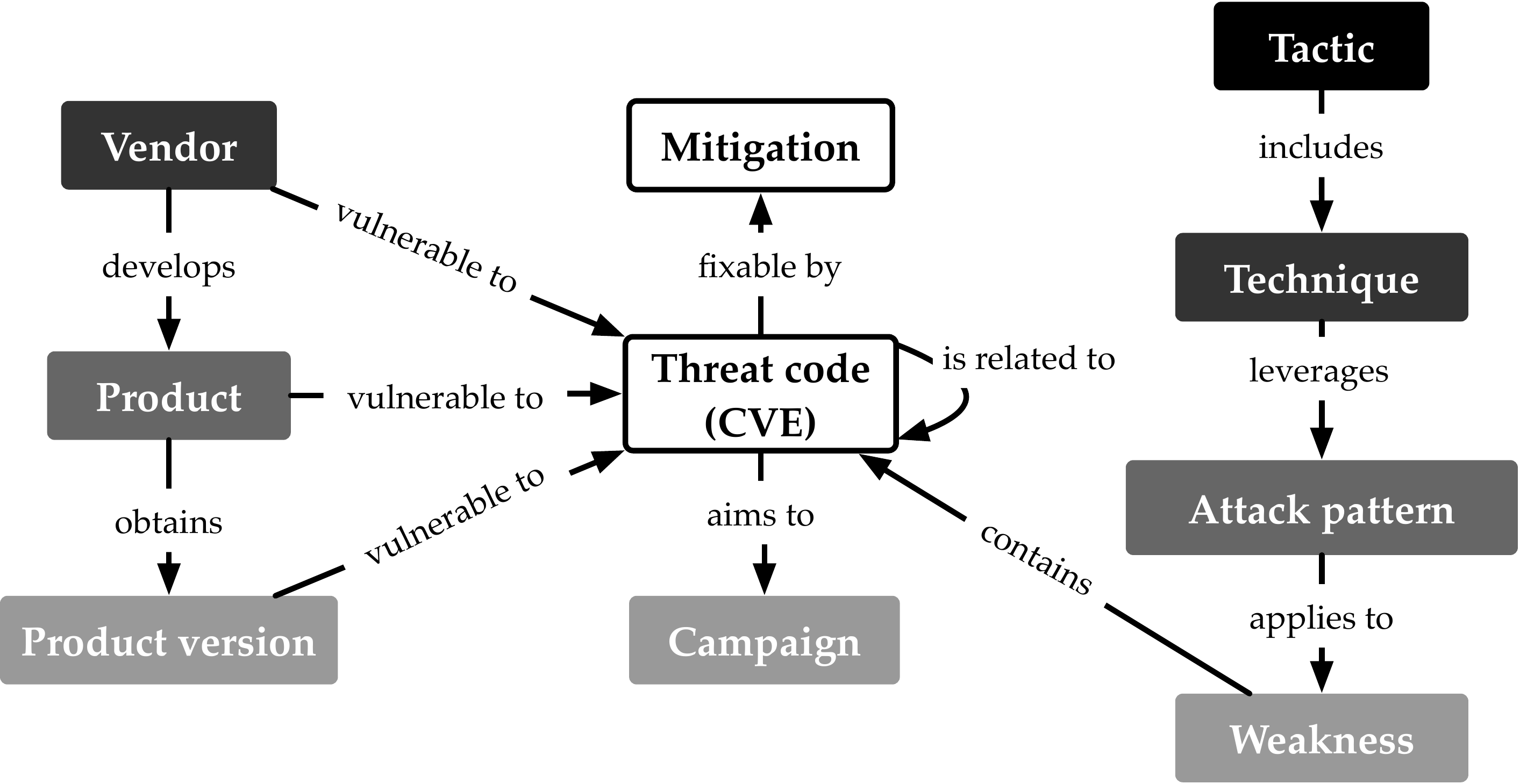, width = 84mm}
    \caption{\footnotesize Structure of our cyber-domain knowledge graph.}
    \label{fig:cyberkg}
\end{figure}

{\bf CyberKG} -- Table\mref{tab:cyber:stat} lists statistics of entities/relations of our cyber-domain \kg, whose structure is shown in Figure\mref{fig:cyberkg}. We add the reverse facts to augment \kg during evaluations.

\begin{figure}[!ht]
    \centering
    \epsfig{file = 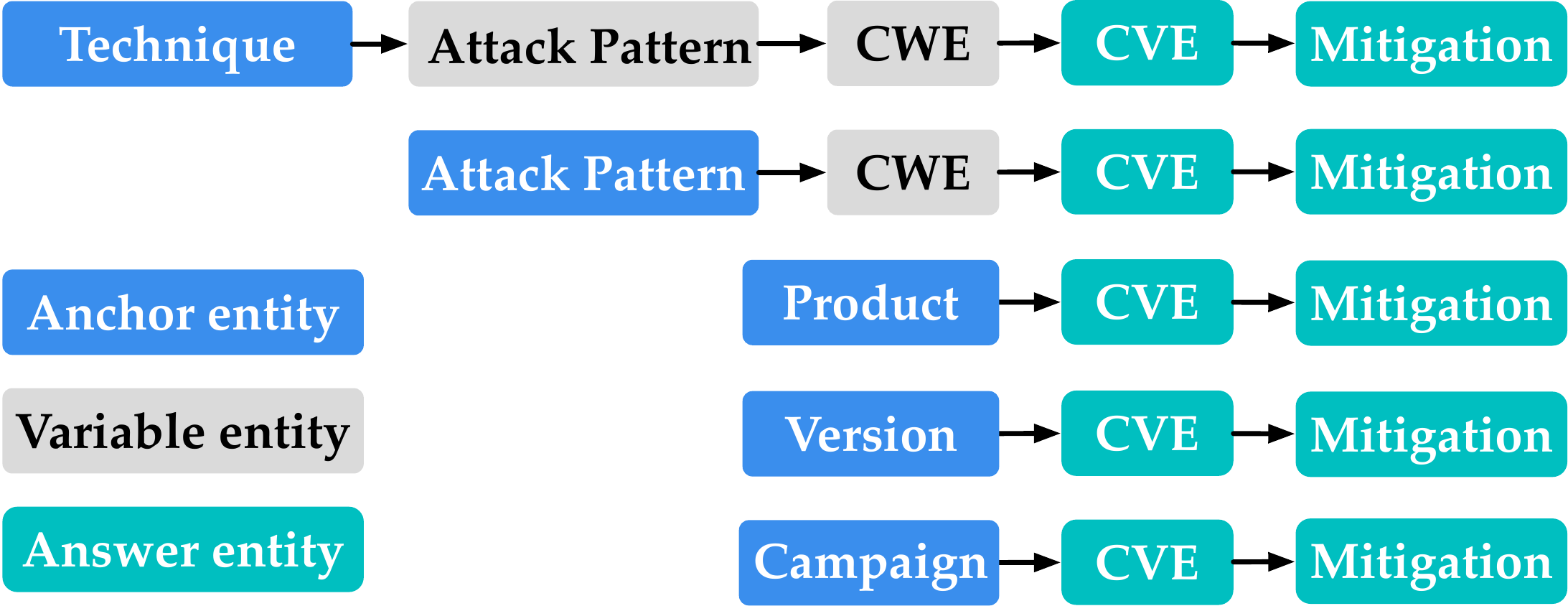, width = 84mm}
    \caption{\footnotesize Different logical paths used in our query set.}
    \label{fig:cyber:qpath}
\end{figure}

\begin{figure}[!ht]
    \centering
    \epsfig{file = 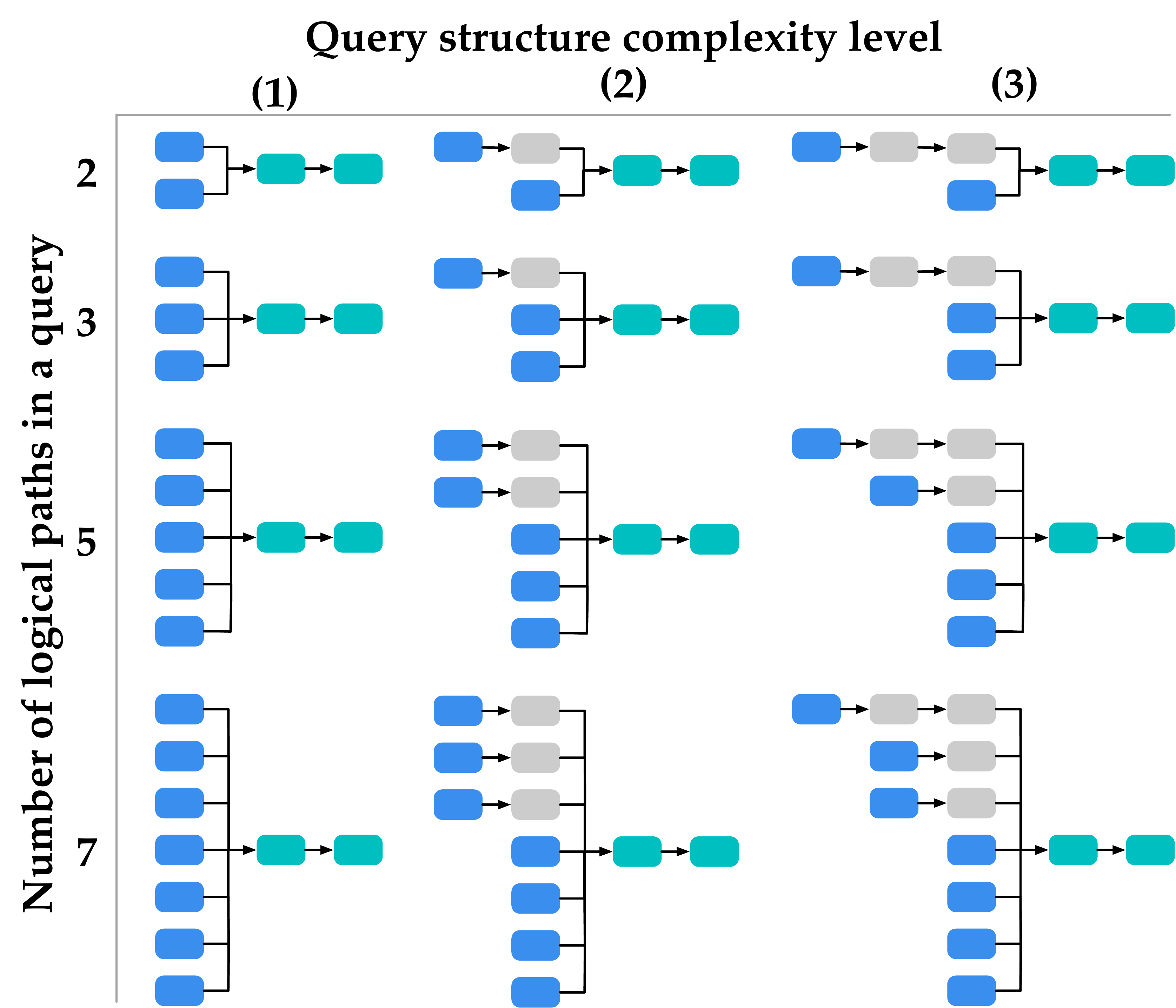, width = 84mm}
    \caption{Topological structures of queries used in cyber-threat hunting (blue -- anchor; grey -- variable; green -- answer).}
    \label{fig:cyber:qstruc}
\end{figure}

Query templates follow Figure \mref{fig:cyber:qstruc}. The first 3 rows (path number=2/3/5) are used as training-query structures, while the last 3 rows (path number = 3/5/7) are used for evaluation. We follow Figure \mref{fig:cyber:qpath} to sample each query path from \kg with a fixed length (1/2/3 hop to \textsf{{\small CVE}} or 2/3/4 hop to \textsf{{\small Mitigation}}), and conjunct the sampled paths together by the \textsf{{\small CVE}} entity. We randomly remove half of the facts from \kg that exists in the evaluation set, which addresses the assumption of incomplete \kg hence relies on the reasoning to find correct answers.

\begin{table}[!ht]{\footnotesize
\centering
\renewcommand{\arraystretch}{1.2}
\begin{tabular}{c|c|c}
{\bf Entity category} & {\bf Data Source} & {\bf Quantity} \\
\hline
{\sf Vulnerability (CVE)} & Webpage, BRON, NVD  & 18,587 \\
{\sf Vendor} & Webpage & 2,223 \\
{\sf Product} & Webpage & 7,103 \\
{\sf Version} & Webpage & 96,725 \\
{\sf Campaign} & Webpage & 13 \\
{\sf Tactic} & BRON &  11\\
{\sf Technique} & BRON & 99 \\
{\sf Attack pattern} & BRON & 323 \\
{\sf Weakness (CWE} & Webpage, BRON & 150 \\
{\sf Mitigation} & NVD & 74,794 \\
\hline
\multicolumn{2}{c|}{{\bf Total}} & {\bf 200,028} \\
\hline
\hline
\multicolumn{2}{c}{\bf Knowledge fact ($v \xrightarrow{r} v'$)} & {\bf Quantity} \\
\hline
\multicolumn{2}{c|}{${\sf Vendor} \xrightarrow{\text{develops}} {\sf Product}$} & 7,897 \\
\multicolumn{2}{c|}{${\sf Product} \xrightarrow{\text{obtains}} {\sf Version}$} & 96,725 \\

\multicolumn{2}{c|}{${\sf Vendor} \xrightarrow{\text{vulnerable to}} {\sf CVE}$} & 26,884   \\
\multicolumn{2}{c|}{${\sf Product} \xrightarrow{\text{vulnerable to}} {\sf CVE}$} & 47,419 \\
\multicolumn{2}{c|}{${\sf Version} \xrightarrow{\text{vulnerable to}} {\sf CVE}$} & 510,781  \\
\multicolumn{2}{c|}{${\sf CVE} \xrightarrow{\text{aims to}} {\sf Campaign}$} & 27,325 \\
\multicolumn{2}{c|}{${\sf CVE} \xrightarrow{\text{is related to}} {\sf CVE}$} &  2502 \\
\multicolumn{2}{c|}{${\sf Tactic} \xrightarrow{\text{includes}} {\sf Technique}$} & 123  \\
\multicolumn{2}{c|}{${\sf Technique} \xrightarrow{\text{leverages}} {\sf Attack pattern}$} & 111  \\
\multicolumn{2}{c|}{${\sf Attack pattern} \xrightarrow{\text{applies to}} {\sf Weakness}$} & 575 \\
\multicolumn{2}{c|}{${\sf Weakness} \xrightarrow{\text{contains}} {\sf CVE}$} & 19,047 \\
\multicolumn{2}{c|}{${\sf CVE } \xrightarrow{\text{fixable by}} {\sf Mitigation}$} &  125,943 \\
\hline
\multicolumn{2}{c|}{{\bf Total (w. reverse facts)}} & {\bf 1,730,664}\\
\hline
\end{tabular}
\caption{Statistics of cyber-domain KG.\label{tab:cyber:stat}}}
\end{table}

{\bf DRKG --} We directly use a constructed drug repurposing \kg. Table \mref{tab:drkg:stat} shows the overall statistics of entity/relation/fact scales; we also add the reverse facts into \textsf{{\small DRKG}}.

\begin{table}[!ht]{\footnotesize
\centering
\renewcommand{\arraystretch}{1.2}
\begin{tabular}{c|c|c}
{\bf \# Entity} & {\bf \# Relation} & {\bf \# Fact (w. reverse)} \\
\hline
97,238 & 107 & 11,748,522 \\
\hline
\end{tabular}
\caption{Statistics of DRKG.\label{tab:drkg:stat}}}
\end{table}

\section{Parameter Setting}
\label{sec:param:setting}

Table\mref{tab:model} summarizes the default parameter setting used in \msec{sec:cyber:case} and \msec{sec:drug:case}.

\begin{table}[!ht]{\footnotesize
\centering
\setlength{\tabcolsep}{4pt}
\renewcommand{\arraystretch}{1.2}
\begin{tabular}{c|r|l}
 Type & Parameter & Setting \\
\hline
\hline
\multicolumn{3}{c}{{\bf Cyber-Threat Hunting}} \\
\hline
\multirow{1}{*}{Encoder $\phi$} & \multirow{1}{*}{Dimension} & 400 \\
\hline
\multirow{3}{*}{Operator $\psi$} & \multirow{2}{*}{ Query2Box Architecture} & 2FC  (Projection) \\
 & & 2FC (Conjunction) \\
 & Hidden Dim & 400 \\
\hline
\multirow{4}{*}{Training} & Learning rate & 0.001 ($\phi$ and $\gamma$)\\
& Batch size & 512 ($\phi$ and $\gamma$) \\
& Epochs & 80000 ($\phi$), 10000 ($\gamma$) \\
& Optimizer & Adam ($\phi$ and $\gamma$)\\
\hline
\multirow{3}{*}{Attack} & $n_\mathrm{kp}$ & 100 \\
& $n_\mathrm{qp}$ & 2 \\
& $n_\mathrm{iter}$ & 5 \\
\hline
\hline
\multicolumn{3}{c}{{\bf Drug Repurposing}} \\
\hline
\multirow{1}{*}{Encoder $\phi$} & \multirow{1}{*}{Dimension} & 200 \\
\hline
\multirow{2}{*}{Operator $\psi$} & \multirow{1}{*}{ R-GCN Architecture} & 2GraphConv \\
 & Hidden Dim & 200 \\
\hline
\multirow{4}{*}{Training} & Learning rate & 0.0001 ($\phi$), 0.001 ($\gamma$)\\
& Batch size & 2048 ($\phi$ and $\gamma$) \\
& Epochs & 100000 ($\phi$), 20000 ($\gamma$) \\
& Optimizer & Adam ($\phi$ and $\gamma$)\\
\hline
\multirow{3}{*}{Attack} & $n_\mathrm{kp}$ & 400 \\
& $n_\mathrm{qp}$ & 20 \\
& $n_\mathrm{iter}$ & 5 \\
\hline
\end{tabular}
\caption{Default parameter setting.\label{tab:model}}}
\end{table}

\section{Additional Results}
\label{sec:additional}  % duplicate

\begin{figure*}[!ht]
    \centering
    \epsfig{file = 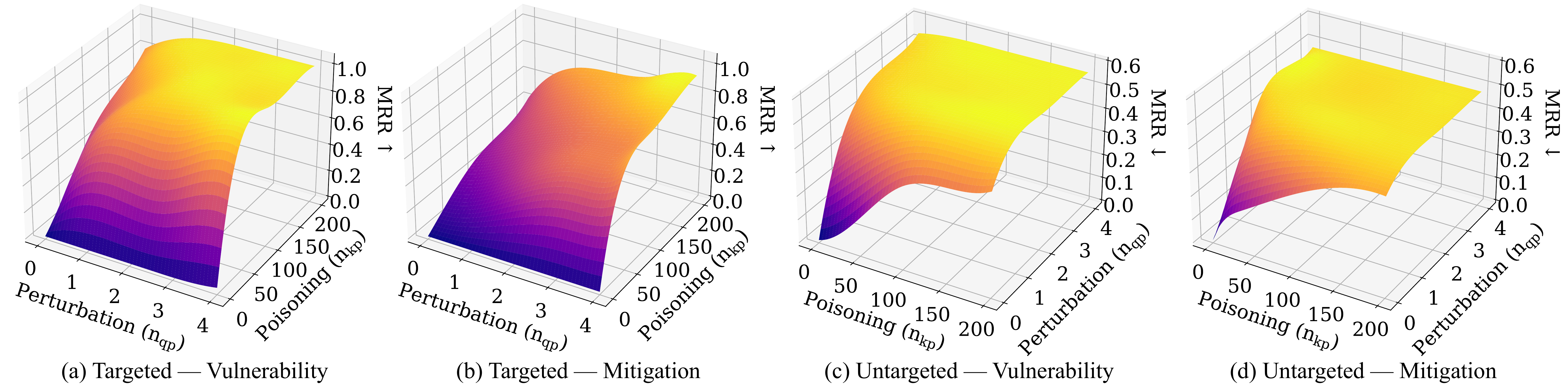, width = 170mm}
    \caption{MRR variation of \aco with respect to knowledge poisoning budget ($n_\mathrm{kp}$) and query perturbation budget ($n_\mathrm{qp}$).}
    \label{fig:cyber:cop:mrr}
\end{figure*}

Figure\mref{fig:cyber:cop:mrr} shows the \mrr variation before and after attacks with respect to $n_\mathrm{kp}$ and $n_\mathrm{qp}$. The observations are similar to Figure \mref{fig:cyber:cop:ndcg}.

\end{document}